\documentclass[12pt]{article}
\usepackage{amsfonts,amsmath,amssymb,amsthm}
\usepackage{scalerel}[2016/12/29]
\usepackage{epsfig,graphicx}
\usepackage{tikz}
\usepackage{float}
\usepackage{braket}
\usepackage{url,hyperref}
\usetikzlibrary{decorations,arrows}
\usetikzlibrary{decorations.markings}
\usetikzlibrary{shapes.geometric}

\newcommand{\hb}{\hskip -0.05cm}
\newcommand{\bb}{\hskip -0.1cm}

\newcommand{\hp}{\hskip -0.05cm + \hskip -0.05cm}
\newcommand{\hm}{\hskip -0.05cm - \hskip -0.05cm}

\newcommand{\he}{\hskip -0.05cm = \hskip -0.05cm}
\newcommand{\sd}{\hskip -0.04cm . \hskip 0.03cm . \hskip 0.03cm . \hskip 0.02cm}

\def\bb{\hskip -0.5mm}
\def\be{\begin{equation}}
\def\ee{\end{equation}}
\def\bea{\begin{eqnarray}}
\def\eea{\end{eqnarray}}
\def\tr{{\rm tr}}

\def\rme{\mathrm{e}}
\def\rmi{\mathrm{i}}

\def\Q{\mathrm{Q}}
\def\rme{\mathrm{e}}
\def\rmi{\mathrm{i}}

\begin{document}

\title{\bf{Combinatorics of generalized Dyck and Motzkin paths}}
\author{Li Gan$^*$, St\'ephane Ouvry$^*$ {\scaleobj{0.9}{\rm and}} \ Alexios P. Polychronakos$^\dagger$}

\date{\today}

\maketitle

\begin{abstract}
We relate the combinatorics of periodic generalized Dyck and Motzkin paths to the cluster coefficients of
particles obeying generalized exclusion statistics, and obtain explicit
expressions for the counting of paths with a fixed number of steps of each kind at each
vertical coordinate. A class of generalized compositions of the integer path length emerges in the analysis.
\end{abstract}

\noindent
* LPTMS, CNRS, Universit\'e Paris-Saclay, 91405 Orsay Cedex, France\\ \noindent {\it li.gan92@gmail.com; stephane.ouvry@u-psud.fr}

\noindent
$\dagger$ Physics Department, the City College of New York, NY 10031, USA and \\ \noindent
The Graduate Center of CUNY, New York, NY 10016, USA
\\ \noindent
{\it apolychronakos@ccny.cuny.edu}

\vfill
\eject
\section{Introduction}
Enumerating closed random walks on various lattices according to their algebraic area amounts to computing traces 
of the ${\bf n}$-th power of quantum Hamiltonians, where ${\bf n}$ is the length of the walks. This approach
was initiated in the study of random walks on the square lattice by relating the problem of their enumeration
to the Hofstadter model of a charged particle hopping on the lattice in the presence of a magnetic field \cite{Hof},
and we shall call such Hamiltonians Hofstadter-like. The generating function of walks weighted by their length
and algebraic area maps to the secular determinant of the Hamiltonian. For specific choices of the parameter dual
to the area, these Hamiltonians can be reduced to finite-size matrices whose near-diagonal structure depends
crucially on the type of walks considered.

Previous work on the subject relied on the computation of the secular determinants of these matrices \cite{Kreft,Shuang}.
Progress in this direction was achieved by interpreting these determinants as grand partition functions
for systems of particles obeying generalized $g$-exclusion quantum statistics in an appropriate 1-body spectrum
\cite{papers} ($g$ is a positive integer, $g=0$ being boson statistics, $g=1$  fermion statistics). Expressing these partition functions in terms of their
corresponding cluster coefficients yields, in turn, the sought for traces. In this process, new combinatorial coefficients 
$c_g(l_1,l_2,\ldots,l_j)$ appear, labeled by the  $g$-compositions of $n=l_1+l_2+\cdots+l_j$ 
with ${\bf {n}}=gn$ ($g=2$ reproducing standard compositions). 
However, a direct combinatorial interpretation of these coefficients was missing.

In this work we bypass the secular determinant and instead tackle directly the trace of the ${\bf n}$-th power
of the  matrices devised to enumerate closed walks on various lattices according to their algebraic area.
We relate the
expression for the trace to periodic generalized Dyck paths (a.k.a.\ {\L}ukasiewicz paths) on a square lattice
with $gn$ horizontal unit steps to the right going vertically either $g\hb-\bb 1$ units up or one unit down
per step and never dipping below vertical coordinate $0$ ($g\hb=\hb 2$ reproducing standard Dyck paths).
Calling ``floor $i$" the level at vertical coordinate $i\hm 1$, we demonstrate that
$gn\,c_g(l_1,l_2,\ldots,l_j)$ counts the number of all possible such paths with $l_1$ up steps from floor 1,
$l_2$ up steps from floor 2, ..., $l_j$ up steps from floor $j$, for a total
of ${\bf n}=gn$ steps inside the strip between floors $1$ and $j\hp g\hm 1$.
In fact, we obtain an even more detailed enumeration of these paths by providing the count of
paths starting with an up step from a given $i$-th floor among the $j+g-1$ floors, and similarly of paths
starting with a down step.

We further extend our results to the enumeration of generalized Motzkin paths that can also move by horizontal
unit steps, by relating such paths to matrices corresponding to mixed $(1,g)$-exclusion statistics 
for particles having either fermionic $g=1$ or $g$-exclusion statistics. The derived expressions for
the corresponding combinatorial coefficients $c_{1,g}$ counting such paths with a fixed number of horizontal,
up, or down steps for each floor are labeled by a further generalized $(1,g)$-composition of the number of steps
$\bf n$. The extension to other classes of paths, corresponding to other generalizations
of quantum exclusion statistics, appears to be within reach of our method.

\section{Square lattice walks: the Hofstadter model}

We start with the original algebraic area enumeration problem for closed walks on a square lattice: among the 
${{\bf n}\choose {\bf n}/2}^2$ closed ${\bf n}$-{steps} walks that one can draw how many of them enclose a given algebraic area $A$? Note that, for closed walks, ${\bf n}$ is necessarily even, ${\bf n}=2n$.

\begin{figure}[H]				
\begin{center}
\includegraphics[scale=0.9]{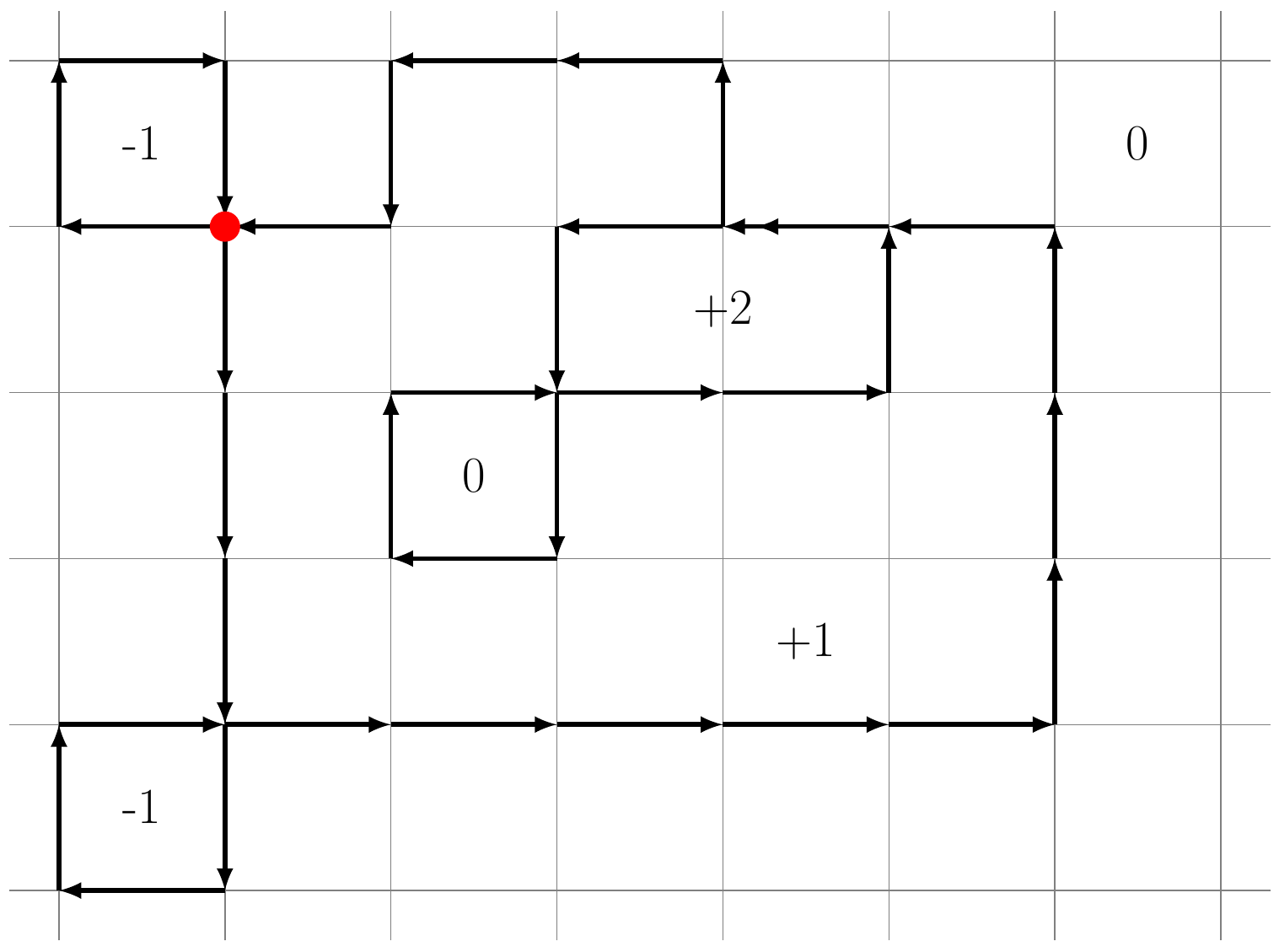}
\end{center}	\vskip -0.2cm
\caption{{\small{A closed walk of length ${\bf n}=36$ starting from and returning to the same bullet (red)
point with winding sectors $m=+2,+1,0,-1$ and various numbers of lattice cells per winding sectors,
respectively $2,14,1,2$. The 
$0$-winding number inside the walk 
arises from a superposition of $+1$ and $-1$ windings. Taking into account the nonzero winding sectors
we end up  with an algebraic area $A=(+2)\times { 2}+ (+1)\times {14}+{(-1)}\times {2}=16$. Note the double arrow on the horizontal link which indicates that the walk has moved 
twice on this link, here in the same left direction.}}}
\label{random walk}
\end{figure}
The algebraic area enclosed by a walk is weighted by its {\it winding numbers}: if the walk moves around a region in 
a counterclockwise (positive) direction its area counts as positive, otherwise negative; if the walk winds around more 
than once, the area is counted with multiplicity (see figure \ref{random walk}). These regions inside the walk are called winding sectors.
Calling $S_m$ the arithmetic area of the $m$-winding sector inside a walk (i.e., the total number of lattice cells it encloses with winding number $m$, where $m$ can be positive or negative) the algebraic area is
\be A=\sum_{m=-\infty}^{\infty}m S_m\;.\nonumber\ee
				
Counting the number of closed walks of length $\bf n$ on the square lattice enclosing an algebraic area $A$ 
can be achieved by introducing two lattice hopping operators $u$ and $v$ in the right and up directions 
obeying
\be v\;u=\Q\; u\;v\nonumber,\ee
and, as a consequence, such that the $u$ and $v$ independent part in 
\be
\big(u+u^{-1}+v+v^{-1}\big)^{\bf n}={\sum_{A} C_{\bf n}(A)\; \Q^A}+\ldots\label{(u+v+u-1+v-1)^n}
\ee
counts the number $C_{\bf n}(A)$ of walks enclosing area $A$. For example, 
$\big(u+u^{-1}+v+v^{-1}\big)^{4}=28 + 4 \Q+4\Q^{-1}+\ldots$ tells that among the ${4\choose 2}^2=36$ 
closed walks of length $4$, $C_{4}(0)=28$ enclose an area $A=0$ and $C_{4}(1)=C_{4}(-1)=4$ 
enclose an {area} $A=\pm 1$. 

Provided that $\Q$ is {interpreted} as $\Q={\rme}^{{\rmi}2\pi\Phi/\Phi_o}$  where $\Phi$ is the flux of
an external magnetic field through the unit lattice cell and $\Phi_o$   the flux quantum,   
\be H=u+u^{-1}+v+v^{-1}\nonumber\ee 
becomes the Hamiltonian for a quantum  particle hopping on a square lattice  and  coupled to a perpendicular  
magnetic field, i.e., the Hofstadter model \cite{Hof}.  Selecting in (\ref{(u+v+u-1+v-1)^n}) the $u,v$ independent part 
of $\big(u+u^{-1}+v+v^{-1}\big)^{\bf n}$ translates in the quantum world to focusing on the trace of $H^{{\bf n}}$ 
with the normalization $\textbf{Tr}\, I=1$, where $I$ is the identity operator.
It follows {that}
\be\textbf{Tr}\, H^{{\bf n}}=\sum_{A} C_{\bf n}(A)\, \Q^{A}\label{eureka},\ee
i.e., the trace gives the generating function of walks weighted by their algebraic area.

When the flux is rational, $\Q={\rme}^{{\rmi}2\pi p/q}$ with $p,q$ coprime integers, 
the lattice operators $u$ and $v$  can be represented by  $q\times q$ matrices
\be \nonumber
u={\rme}^{{\rmi}k_x }\begin{pmatrix}
\Q & 0 & 0 & \cdots & 0 & 0 \\
0 & \Q^{2} &0& \cdots & 0 & 0 \\
0 & 0 & \Q^{3} & \cdots & 0 & 0 \\
\vdots & \vdots & \vdots & \ddots & \vdots & \vdots \\
0 & 0 & 0 & \cdots &\Q^{q-1} & 0 \\
0 & 0 & 0 & \cdots & 0 & \Q^q\\
\end{pmatrix}\;, ~~
v = {\rme}^{{\rmi} k_y}\begin{pmatrix}
\;0\; & \;1\; & \;0\; & \;0\; & \cdots & \;0\; & \;0\; \\
0 & 0 & 1 & 0 &\cdots & 0 & 0 \\
0 & 0 & 0 & 1 &\cdots & 0 & 0\\
\vdots & \vdots & \vdots & \vdots &\ddots & \vdots & \vdots\\
0 & 0 & 0 & 0 &\cdots & 1 & 0\\
0 & 0 & 0 & 0 &\cdots &0& 1\\
1 & 0 & 0 & 0 & \cdots & 0& 0 \\
\end{pmatrix},
\ee
where $k_x$ and $k_y$ are quasimomenta in the $x$ and $y$ directions. It follows that 
$H$ becomes a ${ q\times q }$ matrix as well, and computing {the} trace $\textbf{Tr}\, H^{{\bf n}}$
amounts to taking the matrix trace and integrating over $k_x$ and $k_y$ and dividing by
$(2\pi)^2 q$  for a proper normalization.

\begin{figure}[H]
\begin{center}
\includegraphics[trim=100 405 300 75,clip]{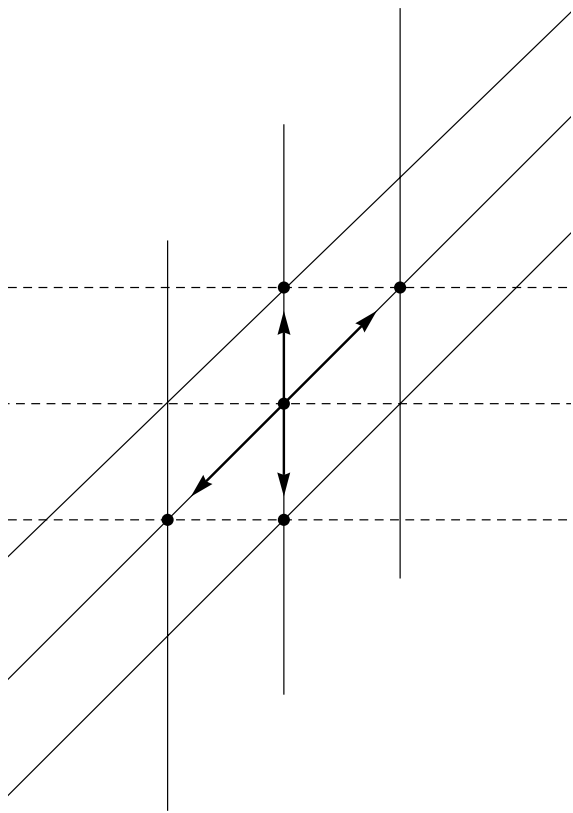}
\end{center}\vskip -0.2cm
\caption{{\small{The deformed square lattice walk steps after the modular transformation.}}}
\label{deformed}
\end{figure}

{One way to evaluate this trace} is to compute the secular determinant  of $H$, namely $\det(I-z H)$. To do so one first performs on $u$ and $v$ the modular transformation 
\be \nonumber
u \to - uv,~~ v \to v,
\ee which preserves the relation $v u = \Q u v$ and the corresponding
traces. It amounts to {looking} at lattice walks on the deformed square lattice of figure \ref{deformed}.

\noindent The Hofstadter matrix becomes 
\be \label{matrix}
H = -uv -v^{-1}u^{-1}+v+v^{-1}= \begin{pmatrix}
{{0}} & {{\omega_{1} }}& 0 & \cdots & 0 & {{\bar{\omega}_{q}}}\\
 {{\bar{\omega}_{1}}} & {{0}} & {{\omega_{2} }}& \cdots & 0 & 0\\
0 & {{\bar{\omega}_{2}}} & {{0}} & \cdots & 0 & 0\\
\vdots & \vdots & \vdots & \ddots & \vdots & \vdots\\
~0~ & ~0~ & ~0~ & \cdots & { {0}} & {{\omega_{q-1} }}\\
\omega_{q} & 0 & 0 & \cdots & {{\bar{\omega}_{q-1}}} & {{0}} \\
\end{pmatrix}
\ee
with $\omega_{k}=(1-\Q^k {\rme}^{{\rmi}k_x}){\rme}^{{\rmi}k_y}$.
Its secular determinant reads
\be \label{g=2 det}
\det(I-z H)=\sum_{n=0}^{\lfloor q/2 \rfloor} (-1)^nZ(n)z^{2n}-2\big(\cos(q k_y)-\cos(q k_x+q k_y)\big)z^q
\ee
with coefficients $Z(n)$ which rewrite as trigonometric multiple nested sums \cite{Kreft}
\be\label{g=2 Zn}
Z(n)= \sum_{k_1=1}^{q-2n+1} \sum_{k_2=1}^{k_1} \cdots \sum_{k_{n}=1}^{k_{n-1}}
s_{k_1+2n-2}
s_{k_2+2n-4} \cdots s_{k_{n-1}+2}
s_{k_{n}},\ee
where $s_{k}=(1-\Q^k)(1-\Q^{-k})=4\sin^2(\pi k p/q)$ (by definition $Z(0)=1$).

From the $Z(n)$'s in (\ref{g=2 Zn}) the algebraic area enumeration can proceed \cite{Shuang,papers} via
\be \log\left(\sum_{n=0}^{\lfloor q/2 \rfloor}Z(n) z^n \right)=\sum_{n=1}^{\infty} b(n) z^n.
\label{name me!}\ee
The $b(n)$'s rewrite as linear combinations of trigonometric simple sums
\be b(n)=(-1)^{n+1}\hskip -0.3cm {\sum_{\substack{l_1, l_2, \ldots, l_{j} \\ { \rm composition}\;{\rm of}\;n}} \hskip -0.4cm {c_2(l_1,l_2,\ldots,l_{j} )} }{\sum _{k=1}^{q-j} s^{l_{j}}_{k+j-1}\cdots s^{l_2}_{k+1} {s}^{l_1}_k},
\label{g=2 b(n)}\ee
where 
\be\label{c2} c_2(l_1,l_2,\ldots,l_{j})= \frac{{l_1+l_2\choose l_1}}{l_1+l_2}\;\; l_2\frac{{l_2+l_3\choose l_2}}{l_2+l_3}\;\cdots \;\; l_{{j}-1}\frac{{l_{{j}-1}+l_{j}\choose l_{{j}-1}}}{l_{{j}-1}+l_{j}}\ee
is labeled by {{the compositions $l_1,l_2,\ldots,l_{j}$ of $n$}}, i.e., the $2^{n-1}$ ordered partitions of $n$;
for example, for $n=3$ one has {the four composition $3=2+1=1+2=1+1+1$}.

\noindent Finally thanks to the identity
$\log\,\det(I-zM)={{\text{tr}}}\,\log(I-zM)$ where ${{\text{tr}}}$ stands for the usual trace of the matrix $M$, we {can show} that the sought after trace reduces to 
\be \textbf{Tr}\, H^{{\bf{n}}=2n}=2n(-1)^{n+1} {1\over q} b(n) \nonumber\ee
so that 
\be
\label{g=2 Tr}\textbf{Tr}\, H^{{\bf n}=2n}=2n\bb{\sum_{\substack{l_1, l_2, \ldots, l_{j} \\ { \rm composition}\;{\rm of}\;n}} \hskip -0.4cm 
{c_2(l_1,l_2,\ldots,l_{j} )}}\,{1\over q}\,{\sum _{k=1}^{q-j} s^{l_{j}}_{k+j-1}\cdots s^{l_2}_{k+1} {s}^{l_1}_k}.
\ee
The trigonometric simple sum $\sum _{k=1}^{q-j} s^{l_{j}}_{k+j-1}\cdots s^{l_2}_{k+1} {s}^{l_1}_k$ in
(\ref{g=2 Tr}) remains to be computed,  which in turn yields the desired algebraic area enumeration via (\ref{eureka}).

Looking at the structure of (\ref{g=2 Zn}) one realizes that, if $s_k$ is interpreted as a spectral function
(Boltzmann factor),
\begin{equation}
s_k = \rme^{ -\beta \epsilon _{k}},
\nonumber
\end{equation}
where $\beta$ is the inverse temperature and $\epsilon_k$ a  1-body spectrum  labeled by an integer $k$,
then  $Z(n)$ is the $n$-body partition function for $n$ particles with 1-body spectrum
$\epsilon_{k}$  and obeying $g=2$ exclusion statistics (no two particles can occupy two adjacent quantum 
states\footnote{For example in the 3-body case one has
\bea \nonumber Z(3)&=&\sum_{k_1=1}^{q-5} \sum_{k_2=1}^{k_1}\sum_{k_3=1}^{k_2}s_{k_1+4}s_{k_2+2}s_{k_3}
\eea
 i.e.,  for $q=7$
\bea Z(3) &=& s_5s_3s_1 + s_6s_3s_1 + s_6s_4s_1 +s_6s_4s_2 =\sum_{{6\ge k_1\ge k_2+2}\;, {k_2\ge k_3+2 }}\; s_{k_1} s_{k_2} s_{k_3}\nonumber,\eea 
 where indeed no adjacent 1-body quantum states contribute. This is the hallmark of 
$g=2$ exclusion with the $+2$ shifts in the nested multiple sums.}) 
and (\ref{g=2 det}) identifies $\det (1- z H)$ as the grand canonical partition function for exclusion-2 particles
in this spectrum with fugacity parameter $-z^2$.
Exclusion statistics is a purely quantum concept which describes the statistical mechanical properties of 
identical particles. Usual particles are either Bosons ($g=0$) or Fermions ($g=1$). Square lattice walks 
invoke statistics $g \hb=\hb 2$, beyond Fermi exclusion. 
In this context, (\ref{name me!}) identifies the $b(n)$'s as nothing but the cluster coefficients of the $Z(n)$'s. 

\section{$g$-exclusion}

We can go a step further by 
 setting the quasimomenta $k_x$ and $k_y$ to zero, since (\ref{g=2 det})
makes it evident that they do not appear in the
$Z(n)$'s. This sets the corners of the Hofstadter matrix (\ref{matrix}) to 
zero\footnote{These particular matrix elements
contribute to spurious umklapp terms in (\ref{g=2 det}) and can be ignored.}, so that  $H$ becomes a particular 
case of the general class of $g=2$ exclusion matrices
\be\label{H2} 
H_2 =\begin{pmatrix}
{0}& {f_{1}} & {0} & \cdots & 0 & 0 \\
{g_{1}} & {0} & {f_{2}} & \cdots& 0 & 0 \\
0 & {g_{2}} & {0} &\cdots & 0 & 0 \\
\vdots & \vdots & \vdots & {\ddots} & \vdots & \vdots \\
0 & 0 & 0 & \cdots & {0} & {f_{q-1}} \\
~0~ & ~0~ & ~0~ & \cdots & {g_{q-1}}& {0} \\
\end{pmatrix},\ee
whose secular determinant ${\det(I-z H_2)}=\sum_{n=0}^{\lfloor q/2 \rfloor} (-1)^nZ(n)z^{2n}$  leads to
the  $Z(n)$'s and the $b(n)$'s in (\ref{g=2 Zn}), (\ref{g=2 b(n)}), (\ref{c2}) and (\ref{g=2 Tr}) with
$ s_k= g_{k}f_{k}$ as spectral function\footnote{The parameter $g$ of $g$-exclusion should not
be confused with the function $g_k$.}. So the enumeration of square lattice walks according to their algebraic area is captured by the $g=2$ exclusion matrix (\ref{H2}), whose hallmark is a vanishing diagonal flanked by two 
nonvanishing subdiagonals $f_{k}$ and $g_{k}$.

The generalization to $g=3$ exclusion leads to the natural matrix form of $H$
\[ H_3=\begin{pmatrix}
{0}& {f_{1}} & 0 &0 & \cdots & 0 &0 &{0} \\
{0} & {0} & {f_{2}} &0&\cdots & 0&0& 0 \\
 {g_{1}} & {0} & {0} &{f_{3}} &\cdots &0 & 0&0 \\
 0&{g_{2}}&{0}&{0}&\cdots &0 & 0&0\\
 \vdots&\vdots & \vdots &\vdots &{\ddots} &\vdots &\vdots & \vdots\\
0 &0 & 0 &0& \cdots &{0} & {f_{q-2}}&0 \\
0 & 0 & 0 &0&\cdots& {0}&{0} &{f_{q-1}} \\
~0~ & ~0~ & ~0~ & ~0~ & \cdots&{g_{q-2}} &{0} & {0} \\
\end{pmatrix}.\nonumber\] 
Now two vanishing diagonal and subdiagonals appear between the $f_{k}$ and $g_{k}$ subdiagonals
(i.e., there is an extra vanishing subdiagonal below the vanishing diagonal).
Computing its secular determinant $\det(I-z H_3)$ yields
\be \nonumber
Z(n) = \sum_{k_1=1}^{q-3n+1} \sum_{k_2=1}^{k_1} \cdots \sum_{k_{n}=1}^{k_{n-1}}
s_{k_1+3n-3}
s_{k_2+3n-6} \cdots s_{k_{n-1}+3}
s_{k_{n}}
\ee
with spectral function $s_k= g_{k}f_{k}f_{k+1}$.  Clearly $Z(n)$ is the partition function
of $n$ particles of exclusion statistics $g=3$ in the one-body spectrum implied by $s_k$.

In general, for $g$-exclusion the Hamiltonian {is}
\be \nonumber
H_g=F(u)v+v^{{1-g}}G(u),
\ee
where $F(u)$ and $G(u)$ are scalar functions of $u$, and amounts to a $g$-exclusion 
matrix\footnote{Indeed the Hofstadter Hamiltonian is a $g=2$ Hamiltonian
\[H=-uv-v^{-1}u^{-1}+v+v^{-1}=(1-u)v+v^{{1-2}}(1-u^{-1}).\]} 
(again ignoring the spurious umklapp matrix elements in the corners)
\be\label{Hg}
H_g=\left(
\begin{array}{ccccccccc}
0 & f_{1} & 0 & \cdots & 0 & 0 & 0 & \cdots & 0 \\
0 & 0 & f_{2} & \cdots & 0 & 0 & 0 & \cdots & 0 \\
\vdots & \vdots & \vdots & \ddots & \vdots & \vdots & \vdots & \ddots & \vdots \\
0 & 0 & 0 & \cdots & 0 & 0 & 0 & \cdots & 0 \\
g_{1} & 0 & 0 & \cdots & 0 & 0 & 0 & \cdots & 0 \\
0 & g_{2} & 0 & \cdots & 0 & 0 & 0 & \cdots & 0 \\
\vdots & \vdots & \vdots & \ddots & \vdots & \vdots & \vdots & \ddots & \vdots \\
~0~ & ~0~ & ~0~ & \cdots & 0 & ~0~ & ~0~ & \cdots & f_{q-1} \\
0 & 0 & 0 & \cdots & g_{q-g+1} & ~~0~~ & ~~0~~ & \cdots & 0 
\end{array}\right),\ee
where now $g-1$ zeros appear between the $f_{k}$ and $g_{k}$ subdiagonals. Its secular determinant
\be \label{g det}
\det(I-z H_g)=\sum_{n=0}^{\lfloor q/g \rfloor} (-1)^nZ(n)z^{gn}\;
\ee
yields
\be \nonumber
Z(n) = \sum_{k_1=1}^{q-gn+1} \sum_{k_2=1}^{k_1} \cdots \sum_{k_{n}=1}^{k_{n-1}}
s_{k_1+gn-g}
s_{k_2+gn-2g} \cdots s_{k_{n-1}+g}
s_{k_{n}}
\ee
with
\be s_{k}= g_{k} f_{k} f_{k+1} \cdots f_{k+g-2}\label{g s_k},\ee
where $k=1,2,\ldots,q\hm g\hp 1$.
Again, as in the $g=2,3$ cases, $Z(n)$ admits an interpretation as the partition function
of $n$ exclusion $g$ particles in the 1-body spectrum implied by the spectral function
$s_k={\rme}^{-\beta\epsilon_k}$, with 1-body levels labeled by the integer $k$.
From  \be\log\left(\sum_{n=0}^{\lfloor q/g \rfloor}Z(n) z^n \right)=\sum_{n=1}^{\infty} b(n) z^n\nonumber\ee
one infers 
\be\label{g bn}
b(n)=(-1)^{n+1}\hskip -0.3cm \sum_{\substack{l_1, l_2, \ldots, l_{j} \\ g{\text{-composition}}\;{\rm of}\;n}} \hskip -0.4cm 
{ {c_g(l_1,l_2,\ldots,l_{j} )} }{{\sum _{k=1}^{q-j-g+2} s^{l_{j}}_{k+j-1}\cdots s^{l_2}_{k+1} {s}^{l_1}_k}}
\ee
with
\be\label{cg}
{c_g (l_1,l_2,\ldots,l_{j})} = 
{{{1 \over l_1}~
\prod_{i=2}^{j} {l_{i-g+1}+\dots +l_i -1 \choose l_i}}}
\ee
with the convention $l_i =0$ if $i\le 0$.
Finally
\be\label{g Tr}\text{tr}\, H_g^{{\bf n}=gn}=gn(-1)^{n+1}b(n)=gn{\sum_{\substack{l_1, l_2, \ldots, l_{j} \\ g{\text{-composition}}\;{\rm of}\;n}} \hskip -0.5cm 
{c_g(l_1,l_2,\ldots,l_{j} )}}{\sum _{k=1}^{q-j-g+2} s^{l_{j}}_{k+j-1}\cdots s^{l_2}_{k+1} {s}^{l_1}_k}.
\ee

These expressions generalize (\ref{g=2 Zn}), (\ref{g=2 b(n)}), (\ref{c2}) and (\ref{g=2 Tr}) to $g$-exclusion 
statistics, {where now} in (\ref{g bn}) {and (\ref{g Tr})} one has to sum over the $g$-compositions of the 
integer $n$, obtained by inserting at will inside the usual compositions (i.e., the $2$-compositions) no more than $g-2$ 
zeros in succession i.e., obtained by allowing up to $g\hm 2$ consecutive integers
in the composition to vanish. For example, one has nine $g=3$-compositions of
					  $n=3$, namely $n=3=2+1=1+2=1+1+1=2+0+1=1+0+2=1+0+1+1=1+1+0+1=1+0+1+0+1$.
				In general  there are $g^{n-1}$ such {$g$-compositions} of the integer $n$ (see \cite{brian}  for an analysis of these extended compositions, also called multicompositions).

\section{Dyck path combinatorics}

We now turn to giving a combinatorial interpretation to the numbers $c_g(l_1,l_2,\ldots,l_{j} )$
in (\ref{cg}), $ l_1,l_2, \dots,l_j$ being a $g$-composition of $n$.
Specifically, we address the question: is there a class of objects whose counting would be determined by
these numbers?

We recall that $c_g(l_1,l_2,\ldots,l_{j})$ was obtained by considering the secular determinant (\ref{g det}) 
and the resulting $n$-body partition functions $Z(n)$ of the $g$-exclusion matrix (\ref{Hg}), and then by 
turning to the associated cluster coefficient $b(n)$ in (\ref{g bn}). On the other hand one sees  that these 
coefficients appear in the trace of the $\bf n$-th power of the $g$-exclusion matrix $H_g$. Let 
us consider directly this trace and denote $h_{ij}$ the matrix elements of $H_g^\mathsf{T}$.
The matrix trace of $H_g^{\bf n}$ becomes
\bea \label{direct}
\tr\,H_g^{\bf n}
&=& \sum_{k_1=1}^q \sum_{k_2=1}^q \cdots \sum_{k_{\bf{n}}=1}^q h_{k_1 k_2}h_{k_2 k_3}\cdots h_{k_{\bf{n}} k_1}.
\eea
The structure of the $g$ exclusion matrix (\ref{Hg}) implies that (\ref{direct}) is a sum of products of 
${\bf n}$ factors $h_{k_i \hskip 0.015cm k_{i+1}}$ with indices such that $k_{i+1}\hm k_i$ take values $g\hm 1$ or $-1$.

We map the sequence of indices $k_1 , k_2 , \dots ,k_{{\bf n}-1}, k_{\bf n}, k_1$
to the heights of a periodic generalized Dyck path of length $\bf n$ starting and ending at height $k_1$,
with vertical steps up by $g\hm 1$ units or down by $1$ unit, denoted as a  [$g\hm 1,-1$] Dyck path
(Figure \ref{Dyck path 3,0,1,1} depicts an example of a $g\he3$ path).
Evaluating the trace (\ref{direct}) amounts to
summing the corresponding products over all such periodic paths, an expression clearly evoking
a path integral. We note that periodic paths must have $n$ up steps and $n(g\hm 1)$ down steps for a total length $gn={\bf n}$.
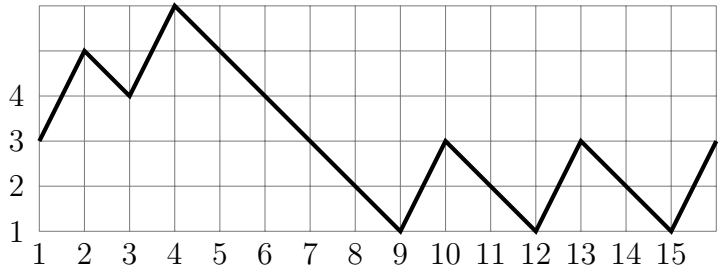
\begin{figure}[H]
\begin{center}
{\begin{tikzpicture}[scale=0.6]
\draw[thin, gray] (0,0) grid (15,5);
\draw[ultra thick, black](0,2)--(1,4)--(2,3)--(3,5)--(8,0)--(9,2)--(11,0)--(12,2)--(14,0)--(15,2);
\node[align=center] at (-0.5,0) {$1$};
\node[align=center] at (-0.5,1) {$2$};
\node[align=center] at (-0.5,2) {$3$};
\node[align=center] at (-0.5,3) {$4$};
\node[align=center] at (0,-0.5) {$1$};
\node[align=center] at (1,-0.5) {$2$};
\node[align=center] at (2,-0.5) {$3$};
\node[align=center] at (3,-0.5) {$4$};
\node[align=center] at (4,-0.5) {$5$};
\node[align=center] at (5,-0.5) {$6$};
\node[align=center] at (6,-0.5) {$7$};
\node[align=center] at (7,-0.5) {$8$};
\node[align=center] at (8,-0.5) {$9$};
\node[align=center] at (9,-0.5) {$10$};
\node[align=center] at (10,-0.5) {$11$};
\node[align=center] at (11,-0.5) {$12$};
\node[align=center] at (12,-0.5) {$13$};
\node[align=center] at (13,-0.5) {$14$};
\node[align=center] at (14,-0.5) {$15$};
\end{tikzpicture}}
\end{center}
\caption{\small{A Dyck path of length $15$ for the $g=3$ composition $3,0,1,1$. The path starts from the third floor with an up step.}}\label{Dyck path 3,0,1,1}
\end{figure}
\vskip -0.35cm
To group together terms with the same weight $h_{k_1 k_2} \cdots h_{k_{\bf n} k_1}$, for each
path we denote by $l_1,l_2,\ldots,l_j$ the number of up steps starting at 1-body level $k , k\hp 1,
\dots, k\hp j\hm 1$ ($k$ is the lowest 1-body level reached by the path). Clearly $l_1+l_2+\cdots+l_j=n$,
and at most $g\hm 2$ successive $l_i$ can vanish, since steps of size $g\hm 1$ can skip $g\hm 2$ levels,
so $l_1 , \dots , l_j$ is
 a $g$-composition of $n$ (figure \ref{Dyck path 3,0,1,1} depicts the $g\he 3$ composition $3,0,1,1$).
Further, each up step $k_i \to k_i\hp g\hm 1$ necessarily implies down steps $k_i\hp g\hm 1 \to k_i\hp g \hm2 ,
\dots k_i\hp 1 \to k_i$, so factors in each term in (\ref{direct}) corresponding to each up step $k_i \to k_i\hp g\hm 1$
contribute the combination
\be
h_{k_i,k_i+g-1} \, h_{k_i+g-1, k_i+g-2}  \cdots h_{k_i+1,k_i}= g_{k_i}\, f_{k_i+g-2} \cdots f_{k_i} = s_{k_i},
\nonumber\ee
where we used (\ref{Hg}) and (\ref{g s_k}). Altogether, the sum in (\ref{direct}) rewrites  as
\be
\tr\,H_g^{\bf n}=\sum_{k=1}^{q-j-g+2} \sum_{\substack{l_1, l_2, \ldots, l_{j} \\ g{\text{-composition}}\;{\rm of}\;n}} \hskip -0.5cm 
C_g(l_1,l_2,\ldots,l_j) s^{l_{j}}_{k+j-1} \cdots s^{l_2}_{k+1} {s}^{l_1}_k,
\nonumber\ee
where $C_g (l_1 , \dots , l_j )$ is the number of periodic generalized Dyck paths of length $gn$ with
$l_1$ up steps originating from the first floor, $l_2$ from the second floor, etc. The sum over $k$
ensures that paths of all starting indices $k_1$ in (\ref{direct}) are included. (Note that the values of 
indices $k,k\hp 1, \dots$ from where up steps can originate map to 1-body levels in the exclusion interpretation.)
Comparing this expression with (\ref{g bn}), we
see that
\be
C_g(l_1,l_2,\ldots,l_j) = gn\, c_g(l_1,l_2,\ldots,l_j).
\nonumber\ee
Therefore, $gn \, c_g(l_1,l_2,\ldots,l_j)$ admits the combinatorial interpretation of the number of
generalized periodic Dyck paths with $l_1 ,\dots,l_j$ up steps from the first, second, etc. floors as defined above.

\subsection{$g=2$}

\begin{figure}[H]
\includegraphics[scale=.7]{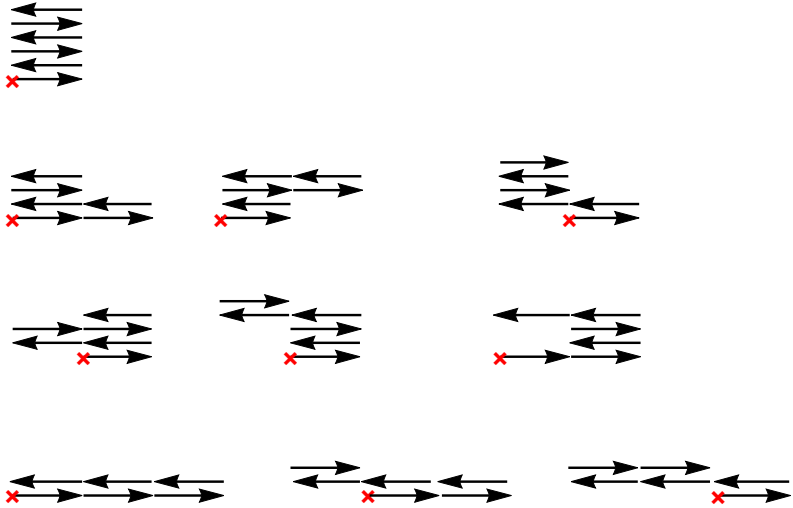}
\caption{\small{The ten closed 1d lattice walks of length ${\bf {n}}\he6$ starting to the right. Their counts are,
from top, $3c_2(3)\he1$, $3 c_2(2,1)\he3$, $3c_2(1,2)\he3$ and $ 3 c_2 (1,1,1)\he3$. The four sets of walks
correspond to the four compositions of $n\he3$,
namely $3\he2+1\he1+2\he1+1+1$.  The walks have been spread in the vertical direction for clarity,
and to demonstrate their correspondence with Dyck paths.
Red crosses denote the starting and ending point of each walk.}}
\label{nis6}
\end{figure}

\begin{figure}[H]
\includegraphics{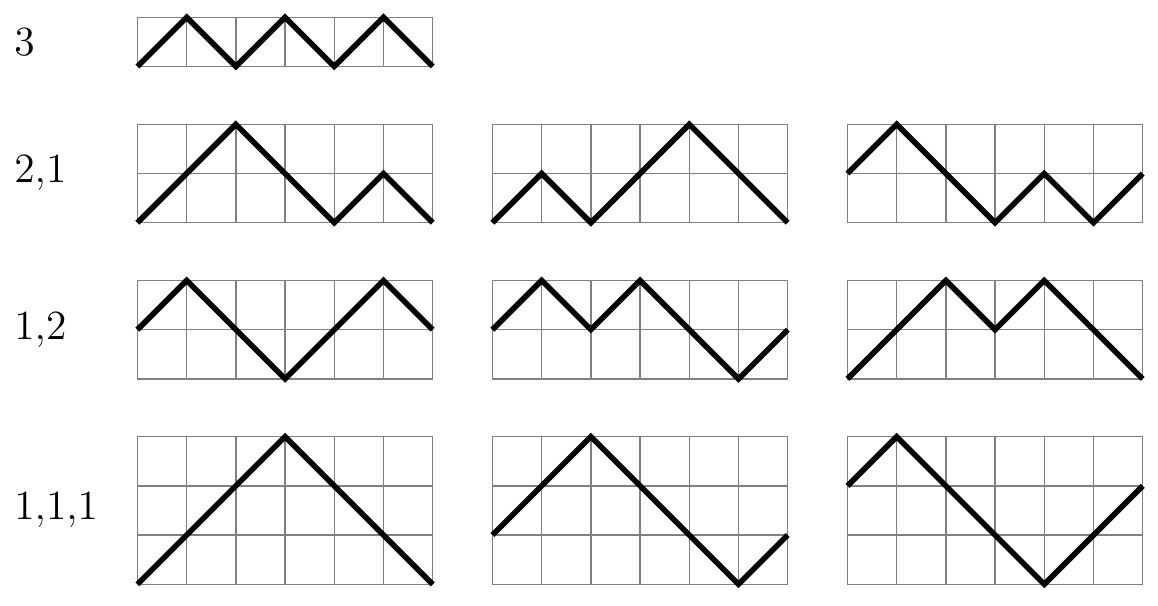}
\caption{\small{The ten Dyck paths of length ${\bf {n}}=6$ starting with an up step.
They are in one-to-one correspondence with the 1d walks of figure \ref{nis6}.}}
\label{dyckfig}
\end{figure}

We focus on the simplest nontrivial case $g=2$ and derive the combinatorics.
The combinatorial interpretation of $c_2 (l_1,l_2,\ldots,l_{j} )$ was already hinted at in \cite{Shuang}, where
it was remarked that $n c_2(l_1,l_2,\ldots,l_{j} )$ counts the number of closed random walks of length
${\bf {n}}=2n$ on a 1d lattice starting towards the right, containing $l_1$ right-left steps on top of each other
followed by $l_2$ right–left steps on top of each other, etc. as shown for ${\bf {n}}=6$ in figure \ref{nis6}. 
It is easy to see that such closed walks
map to periodic Dyck paths starting with an up step, by rotating
the lattice by $\pi/2$ and performing one horizontal step to the right with each walk step, as in figure \ref{dyckfig}.
We conclude that $n c_2(l_1,l_2,\ldots,l_j)$ counts the total number of periodic Dyck paths of length ${\bf n}=2n$
starting with an up step and having $l_1$ up steps originating from the first floor, $l_2$ from the
second floor, etc. The remaining count $n c_2(l_1,l_2,\ldots,l_j)$ corresponds to paths starting with a down step,
since for $g=2$ the two sets of paths map to each other through reflection with respect to the horizontal,
for a total of  $2n c_2(l_1,l_2,\ldots,l_j)$ paths.

We can also infer the more granular Dyck path counting that
\be \label{l_i c_2}
l_i\, c_2(l_1,l_2,\ldots,l_j)=\prod_{k=1}^{i-1}{l_{k}+l_{k+1}-1\choose l_{k}}~\prod_{k=i}^{j-1}{l_{k}+l_{k+1}-1\choose l_{k+1}}
\ee 
counts the number of periodic Dyck paths of length ${\bf n}=2n$ starting from the $i$-th floor with an up step and
having $l_1$ up steps originating from the first floor, $l_2$ steps from the second floor, etc.
(The result $l_1\,\hb c_2(l_1,l_2,\ldots,l_j)$ for $i=1$ was also derived in \cite{Krattenthaler, italian paper}.)
Clearly the sum of the above counts for all $i=1,2,\dots,j$ reproduces the total count of starting up
paths $n\, c_2(l_1,l_2,\ldots,l_j)$.

\begin{figure}[H]
\begin{center}
\includegraphics{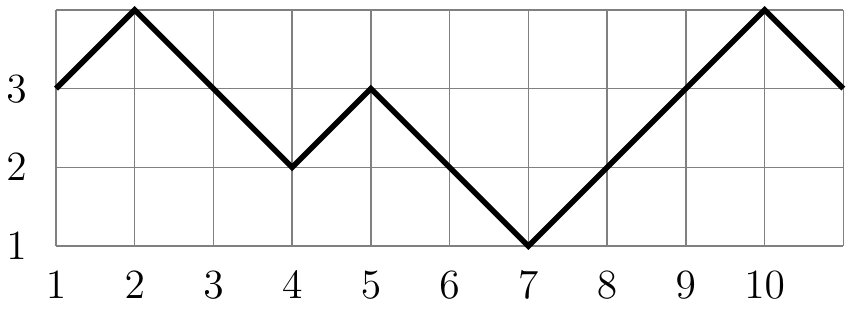}
\caption{\small {A periodic Dyck path of length $10$ starting at $i=3$, characterized by the sequence
$i_s=3,2,1,2,3$ of floors from which the up steps start. One can uniquely reconstruct the path from this 
sequence by producing the sequence of up step positions $p_s = i\hm i_s \hp 2s \hm 1 \he 1, 4, 7, 8, 9$ and filling
the gaps with down steps at positions $2, 3, 5, 6, 10$.
}}\label{Dyck path of length 10}
\end{center}
\end{figure}

To give a proof of  (\ref{l_i c_2}), we remark that the sequence of floors where an up step starts
fully determines the path. This is obvious since there is a unique way to ``fill in" the
remaining down steps to form a path, and can be made explicit by the relation
\be
p_s = i - i_s + 2s-1 ~,~~~ s=1,\dots, n
\nonumber\ee
where $p_s \in[ 1, 2n]$ is the step of the path at which the $s^\text{th}$ up step occurs
and $i_s\in [1,j]$ is the floor at which it occurs. $(p_s , i_s )$ thus determine both the position and floor at
which each up step occurs, fully fixing the path. This is illustrated in figure \ref{Dyck path of length 10} for a periodic path of length 10.
It follows that, to enumerate all possible periodic Dyck paths starting from a given
floor $i$ with an up step, it is sufficient to count all the possible sequences of floors where
an up step starts, given the constraint that $l_1$ up steps are on the first floor, $l_2$ steps
on the second floor, etc. Note, however, that admissible floor sequences satisfy the additional constraint
$i_{s+1} \le i_s \hp 1$ (arising from $p_{s+1} > p_s$) as well as the starting condition at floor $i$, namely  $i_1 = i$.
\begin{figure}[H]
\includegraphics[scale=0.43]{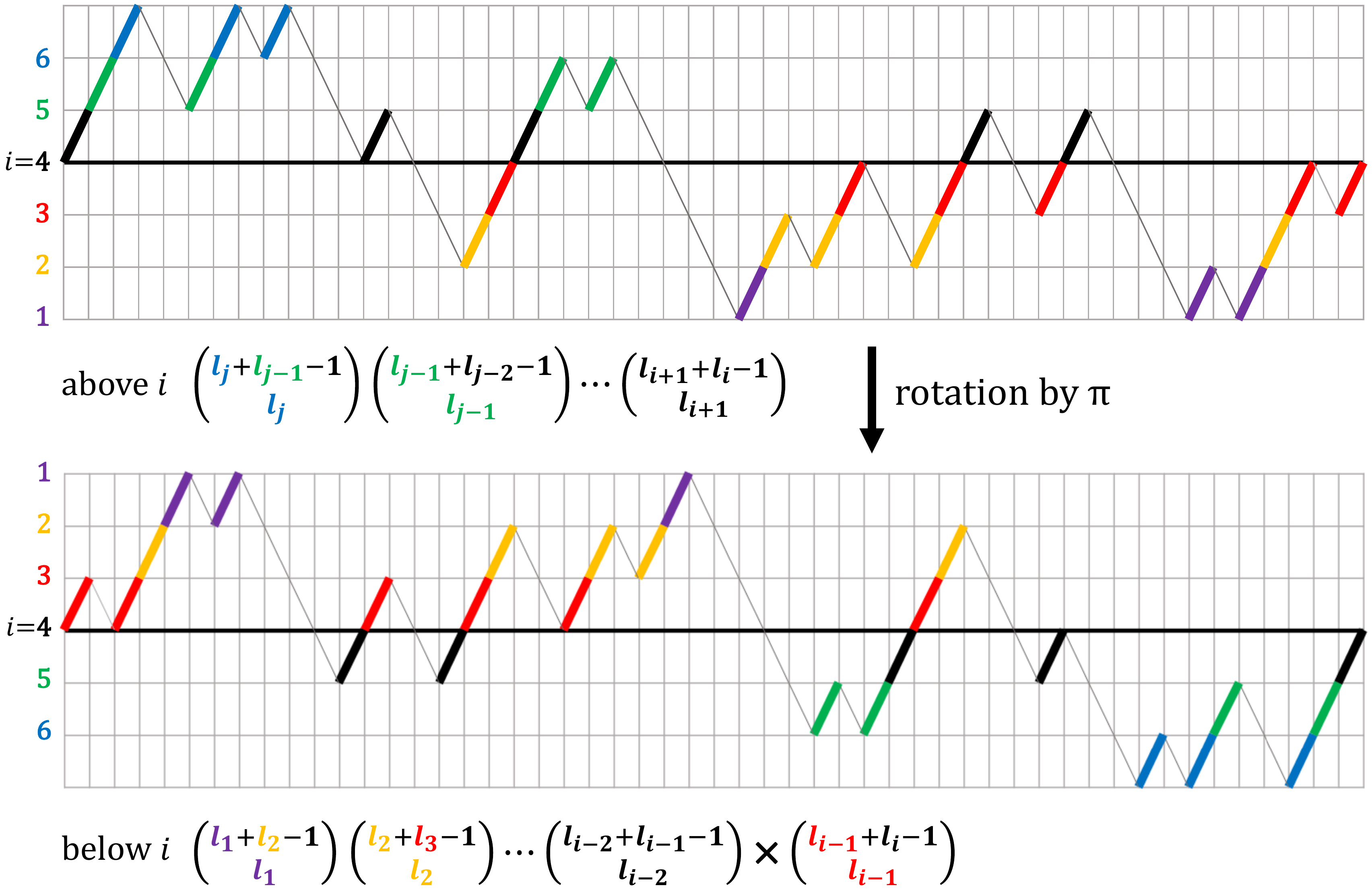}
\caption{\footnotesize{A Dyck path starting and ending at floor $i=4$. The first (blue) up step on floor 6
is connected to the (green) up step on floor 5 at its left, and the next two floor 6 up steps are connected
to each other and the floor 5 up step to their left. The steps on floor 5 become ``units" with the floor 6
steps attached to them, and are attached to floor 4 (black) up steps to their left. Rotating by $\pi$,
(purple) up steps on floor 1 are connected to (orange) up steps on floor 2 to their left, and similarly
for steps on floors 2 and 3 (red). The ordering of up steps on the remaining floors 4 and 3
(except the starting step) is unrestricted.}}
\label{dyckpathc2}
\end{figure}
To count these configurations efficiently, we start from the top two floors $j$ and $j\hm 1$. Since these floors
are above the floor $i$ of the starting step, the first up step among the $l_{j-1}$ and $l_j$ steps on these
floors must necessarily be on the $j\hm 1$ floor. Now notice that all floor $j$ up steps that are between any
two floor $j\hm1$ steps {\it must be connected to the left floor $j\hm 1$ up step and to each other}
(see figure \ref{dyckpathc2}). Therefore, the configuration of floor $j$ steps is fully fixed by the floor
$j\hm 1$ steps and by the distribution of
$j$ floor steps between them. The number of different ways the $l_j$ steps can be distributed among the
remaining $l_{j-1} \hm1$ steps after the first one is $l_{j-1} + l_j -1 \choose l_j$.

Moving to the next two floors, $j\hm 1$ and $j\hm 2$, we can repeat the above argument between the
$l_{j-2}$
and $l_{j-1}$ up steps. Each floor $j\hm 1$ up step comes with a fixed set of floor $j$ up steps
attached and constitutes one compact unit. The distribution between the $l_{j-2}$ steps and the $l_{j-1}$ units,
with the first step again on the $j\hm 2$ floor, fully fixes the positions of the $l_{j-1}$ units, and there are
$l_{j-2} + l_{j-1} -1 \choose l_{j-1}$ such configurations. The argument can be repeated as long as all steps
are above $i$, that is, down to floors $i$ and $i\hp 1$, giving an overall multiplicity of paths with fixed up steps
on floors $i$, $i\hm1$, $\dots , 1$
\be
C_{\text {above}\,i}=\prod_{k=i}^{j-1} {l_k + l_{k+1} -1 \choose l_{k+1}}.
\nonumber\ee
Once we dip below $i$ the situation changes. For floors  $k$ and $k\hm 1$, $k\le i$, the first up step could be
either on floor $k$ or on $k\hm 1$, and the configuration of floor $k$ units in {\it not} fixed by the floor $k$
up steps. To deal with floors below $i$, we rotate the path by $\pi$, which inverts floors as well as the direction
of the path but leaves up steps as up steps. The bottom floors $1$ and $2$ now effectively become top floors, and
the situation is similar to floors $j$ and $j\hm 1$. The first up step is necessarily at floor 2, and
a similar argument as before gives the multiplicity of paths with fixed up steps on floor 2 as
$l_1 + l_2 -1 \choose l_1$. Repeating the above argument for higher floors, as long as all steps are
below $i$, that is, up to floors $i\hm 2$ and $i\hm 1$, we get an
overall multiplicity of paths with fixed up steps on floors $i\hm 1$, $i$, $\dots , j$
\be
C_{\text{below}\,i}=\prod_{k=1}^{i-2} {l_k + l_{k+1} -1 \choose l_k}.
\nonumber\ee
Note that we cannot extend the argument to floors $i\hm 1$ and $i$ since up steps originating at floor $i$
lie {\it above} $i$ (and thus, upon inversion, below it).

The product of the above two factors gives the multiplicity of paths with fixed up steps on floors $i\hm 1$
and $i$ (the only two left fixed by above-$i$ or below-$i$ considerations). The full multiplicity of paths
can be determined by also considering the relative placement of the $l_{i-1}$ and $l_i$ up steps on these two
floors. They can, in principle, be distributed at will, except that we have fixed the first up step to be on floor $i$.
The remaining $l_i \hm 1$ and $l_{i-1}$ steps can be distributed in ${l_{i-1} + l_i -1 \choose l_{i-1}}$ ways,
which contributes the missing factor $k=i\hm 1$ in the product for $C_{\text {below }i}$.
Combining the factors reproduces (\ref{l_i c_2}).

Note that if we relax the condition that paths start with an up step from the starting floor $i$, then all
$l_i$ and $l_{i-1}$ steps can be distributed at will and contribute a multiplicity ${l_{i-1} + l_i \choose l_{i-1}}$.
This differs by a factor $(l_{i-1} + l_i )/l_i$ from the previous result and gives the result
\be
C_i = (l_i + l_{i-1}) \,c_2 (l_1 , l_2 , \dots, l_j )
\nonumber\ee
for the number of paths starting and ending at floor $i$.  (This result  was also derived in \cite{italian paper}.)
Summing over $i= 1,2,\dots,j\hp 1$ (with $l_0 = l_{j+1} = 0$) reproduces the full number of paths 
$2n \, c_2 (l_1 , l_2 , \dots, l_j )$.

We also note that the case of paths starting with a down step at floor $i=2,\dots,j+1$ can be dealt with
in a similar way, with the difference that now the first up step on floors $i\hm 1$ and $i$
happens at floor $i\hm 1$, so their relative arrangement has a multiplicity $l_{i-1}+l_i -1\choose l_i$,
which yields, as expected, the result $l_{i-1}\, c_2 (l_1 , l_2 , \dots, l_j )$. This can also be obtained graphically by  
\vskip -0.1cm

$\bullet$
cutting the periodic Dyck paths starting with an up step from the $(i\hm1)$-th floor at the last occurrence of a
down step from the $i$-th floor

\begin{figure}[H]
\includegraphics{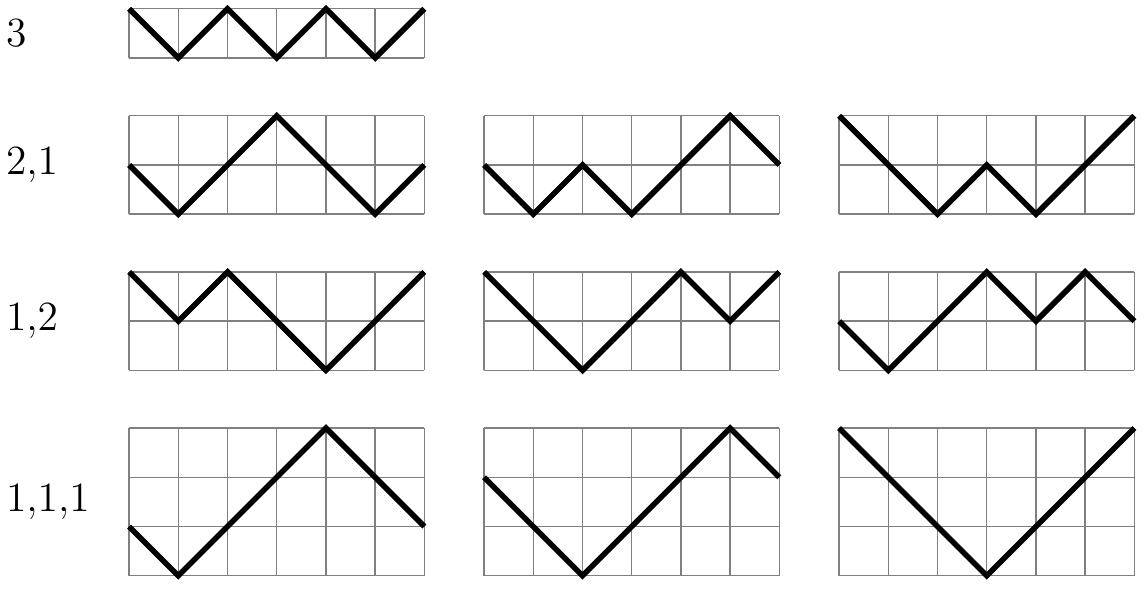}
\caption{\small{The ten Dyck paths of length ${\bf {n}}=6$ starting with a down step. Their counts are $3c_2(3)=1$, $3 c_2(2,1)=3$, $3c_2(1,2)=3$ and $ 3 c_2 (1,1,1)=3$.}}
\label{dyckfig2}
\end{figure}

\vskip -0.1cm
$\bullet$ interchanging the two pieces (see figure \ref{dyckfig2}).
\noindent This establishes a one-to-one mapping between paths starting with an up step from floor $i\hm 1$ and
paths starting with a down step from floor $i$ and shows that $ l_{i-1} \, c_2(l_1,l_2,\ldots,l_j)$ also counts the
number of these paths, consistent with the analytical result as well as the counting
$(l_i + l_{i-1}) \, c_2 (l_1 , l_2 , \dots, l_j )$
for (up- or down-starting) periodic paths starting at floor $i$ derived before, and the total number
$2n\, c_2(l_1,l_2,\ldots,l_j)$ of such periodic paths.



\subsection{General $g$}\label{Geng}

These results can be generalized to $g$-exclusion paths. Consider periodic generalized $[g-1,-1]$ Dyck paths of length
${\bf n}=gn$ with $n$ up steps, each going up $g\hm1$ floors, and $(g\hm1)n$ down steps, each going
down 1 floor, and thus confined between the 1st and ($j\hp g\hm 1$)-th floor. The number of paths starting
with an up step from the $i$-th floor is 

\vskip -0.8cm
{{\footnotesize{
\bea \label{g two}
\hskip -0.6cm l_i\, c_g(l_1,l_2,\ldots,l_j)&=&{\bb\frac{(l_{i-g+1}+\cdots+l_i \hm 1)!}{l_{i-g+1}! \cdots l_{i-1}! \,(l_i\hm 1)! }
\prod_{k=1}^{i-g}}{l_{k}\hp\cdots\hp l_{k+g-1}\hm 1\choose l_{k}}
\prod_{k=i-g+2}^{j-g+1}\bb\bb{l_{k}\hp\cdots\hp l_{k+g-1}\hm 1\choose l_{k+g-1}}\;\\
&=&\frac{(l_i+\cdots+l_{i+g-2}\hm 1)!}{(l_i\hm 1)!~l_{i+1}!~\cdots l_{i+g-2}!}~\prod_{k=1}^{i-1}
{l_{k}\hp\cdots\hp l_{k+g-1}\hm 1\choose l_{k}}~\prod_{k=i}^{j-g+1}
{l_{k}\hp\cdots\hp l_{k+g-1}\hm 1\choose l_{k+g-1}}
\nonumber\eea}}}

\vskip -0.3cm
\noindent
with the conventions $l_i=0$ for $i<1$ or $i>j$, and $\prod_{k=m}^n (\,\cdots)=1$ for $n<m$ understood.

The proof of this formula can be achieved with a method similar to the one for $g=2$, appropriately
generalized. Again, the sequence
$i_s$ of floors with up steps fixes the path, the positions $p_s$ of up steps in the path being given by
\be\nonumber
p_s = i - i_s  + g (s-1) +1.
\ee
Similarly to the $g=2$ case, the up steps on the top floor $j$ are fully determined by their relative
position with respect to the up steps on the $g\hm 1$ floors below it $j\hm 1 , \dots, j-g+1$
(which can only be followed by down steps), with the first up
step always occurring in one of these lower floors, for a multiplicity of $l_{j-g+1} + \cdots + l_j -1 \choose l_j$. 
Up steps on floors $j\hm g \hp 1, \dots, {j\hm 1}$ now constitute units with any possible floor $j$ steps
attached to them, and the argument can be repeated for all successive sets of $g$ floors
$k, k\hm 1, \dots , k\hm g \hp 1$, accumulating multiplicity factors $l_{k-g+1} + \dots + l_k -1\choose l_k$
down to $k=i\hp 1$.

The same argument can be used for steps below $i$ after a $\pi$ rotation, starting from the
bottom $g$ floors $1, \dots, g$ and for all higher floors $k, k\hp 1 , \dots, k \hp g \hm 1$, accumulating
multiplicity factors $l_k + \dots + l_{k+g-1}- 1 \choose l_k$ up to $k=i\hm g$ (up steps that connect to floor
$k$ below $i$ {\it downwards}, as the rotation by $\pi$ argument requires, originate from floor $k\hp g \hm 1$,
and $k = i\hm g$ is the highest floor for which this step originates below floor $i$).

The multiplicities picked up for steps below and above floor $i$ reproduce the products in the upper form of
({\ref{g two}). The combination of the two reduction processes leaves a common set of fixed steps on floors
$i\hm g \hp 1 , \dots, i$. The relative placement of these steps can be chosen at will,
with the constraint that the first up step is from floor $i$, giving a multiplicity of
\be\nonumber
{l_{i-g+1} + \cdots + l_i -1 \choose l_{i-g+1} ~, \dots ,~ l_{i-1}} =
\frac{(l_{i-g+1}+\cdots+l_i \hm 1)!}{l_{i-g+1}! \cdots l_{i-1}! \,(l_i\hm 1)! }
\ee
reproducing the remaining factor in the first line of (\ref{g two}).
For $i \le g$ there are no steps below $i$ to consider and the reduction above $i$ may need to terminate
at a floor higher than $i\hp 1$, and for $j \le g$ all up steps can arise in any arrangement, and these cases
are captured by the conventions below (\ref{g two}). The rewriting in the second line minimizes the
use of these conventions.

The number of paths starting from floor $i$ with either an up or down step can also be derived.
In this case, there is no requirement that the first step among floors $i\hm g \hp 1 , \dots, i$ must be on
floor $i$, and the relative placement of up steps is unrestricted. Their multiplicity is
${l_{i-g+1} + \cdots + l_i \choose l_{i-g+1} , \dots , l_{i-1}}$, which gives the result for the number of
paths $(l_{i-g+1} + \cdots + l_i ) \, c_g (l_1 ,\dots, l_j)$. Correspondingly, the number of paths starting
with a down step from floor $i$ is deduced by subtracting $l_i \,  c_g (l_1 ,\dots, l_j)$ from the full count,
yielding $(l_{i-g+1} + \cdots + l_{i-1}) \, c_g (l_1 ,\dots, l_j)$, and the total number of paths is obtained by
summing over all $i$ as $gn\hskip 0.05cm c_g (l_1 ,\dots, l_j)$.

The derivation of the counting formula can be substantially simplified using the following alternative
approach based on cyclic permutations that bypasses the subtleties around the starting floor $i$.

We first calculate the number of paths starting from the lowest floor $i=1$. Now all steps originate above $i$,
and the reduction argument applies down to floors $1, \dots , g$, giving the multiplicity of paths
for a fixed set of up steps on floors $1$ through $g\hm 1$ as the
product of factors $l_k + \cdots + l_{k+g-1}-1 \choose l_{k+g-1}$ for $k=1, \dots, j\hm g \hp 1$.
The placement of up steps  that start in the   bottom $g\hm 1$ floors is arbitrary with the exception that the first up step
occurs on floor $1$, for a multiplicity of $l_1 + \cdots + l_{g-1} -1 \choose l_2 , \dots , l_{g-1}$. Altogether,
the number of paths starting at the bottom floor is
\be
{(l_1 + \dots + l_{g-1}-1 )! \over (l_1 -1)! l_2! \cdots l_{g-1}!} \prod_{k=1}^{j-g+1}{l_{k}+\cdots+
l_{k+g-1}\hm 1\choose l_{k+g-1}} = l_1\, c_g (l_1 , \dots , l_j ).
\ee
The number of all possible paths can be obtained by circularly permuting the $gn$ steps of paths starting at the bottom, which produces $gn\, l_1\, c_g (l_1,\ldots,l_j)$ paths. However, each time an up step from the
1st floor occurs first, it reproduces the set of paths starting at the bottom. Since there are $l_1$ such steps,
this results in an overcounting by a factor $l_1$. Correcting for this, we recover the total number of paths as
$gn \, c_g (l_1,l_2,\ldots,l_j)$.

The count of paths starting with an up step at floor $i$ can be obtained with a similar argument.
Cyclically permuting these paths reproduces, again, all possible paths, but with an overcounting by a factor of
$l_i$, since each time that an up step at floor $i$ becomes first it reproduces the full set. Therefore,
we obtain a count of $l_i \, c_g (l_1 , \dots , l_j )$ as obtained before.
The number of paths starting with a down step from the $i$-th floor, with $i=2,3,\ldots,j\hp g\hm 1$, can
also be reproduced graphically:

$\bullet$ consider all periodic paths starting with an up step from either the ($i\hm 1$)-th or the  ($i\hm 2$)-th,
or $\ldots$ the $(i\hm g\hp 1)$-th floor, and cut them at the last occurrence of a down step from the $i$-th floor

$\bullet$ interchange the two pieces.

\noindent
This, again, establishes a one-to-one correspondence between the two sets of paths, and gives the
number of paths starting with a down step from floor $i=2,\dots, j\hp g\hm 1$ and $l_1, l_2 , \dots ,l_j$
up steps from each floor as $(l_{i-g+1} + \cdots + l_{i-1}) \, c_g (l_1 ,\dots, l_j)$, as obtained before.


\section{A generalization: $(1,g)$-exclusion statistics and generalized Motzkin paths}

\subsection{$(1,2)$-exclusion statistics}

In \cite{honeycomb} we tackled the algebraic area enumeration of closed walks on a honeycomb lattice.
Again a Hofstadter-like Hamiltonian was central to the enumeration, rewritten as  a $2q\times 2q$
matrix, which was subsequently reduced to a  $q \times q$  matrix. 
The essence of the enumeration was encapsulated in the  exclusion matrix 
\be \nonumber
H_{1,2}=\begin{pmatrix}
\tilde{s}_1 & f_{1} & 0 & \cdots & 0 & 0\\
g_{1} & \tilde{s}_2 & f_{2} & \cdots & 0 & 0 \\
0 & g_{2} & \tilde{s}_3 & \cdots & 0 & 0 \\
\vdots & \vdots & \vdots & \ddots & \vdots & \vdots \\
0 & 0 & 0 & \cdots & \tilde{s}_{q-1} & f_{q-1} \\
~0~ & ~0~ & ~0~ & \cdots & g_{q-1} & \tilde{s}_q \\
\end{pmatrix}.
\ee
In addition to the two subdiagonals $f_{k}$ and $g_{k}$, a hallmark of $g=2$ exclusion,
$H_{1,2}$ also has a nonvanishing $\tilde{s}_k$ main diagonal, a hallmark of $g=1$ statistics, i.e., Fermi statistics,  as it indeed
describes particles obeying a mixture of the two statistics $g=1$ and $g=2$.

The secular determinant reads
\be \label{1,2 det}
\det(I-z H_{1,2})=\sum_{n=0}^{q} (-z)^n Z(n),
\ee
where $Z(n)$ can be interpreted as the $n$-body partition function for particles in a 1-body spectrum 
$\epsilon_1,\epsilon_2,\ldots,\epsilon_k,\ldots, \epsilon_q$ with fermions occupying 1-body energy level $k$ with Boltzmann
factor ${\rme}^{-\beta \epsilon_k} = \tilde{s}_k$ and two-fermion bound states occupying  1-body energy levels $k,k+1$ with
Boltzmann factor ${\rme}^{-\beta \epsilon_{k,k+1}} = -g_{k}f_{k} := -s_k$\footnote{In the pure $g=2$ case
we took the Boltzmann factors of exclusion particles (bound states) as $+s_k$ and compensated by
absorbing the negative sign in the fugacity $-z^2$. In the mixed $1,g$ case we have no such flexibility,
although the alternative, more symmetric choice ${\rme}^{-\beta \epsilon_k} = -\tilde{s}_k$,
${\rme}^{-\beta \epsilon_{k,k+1}} = -s_k$ and fugacity $+z$ could have been made.}. Since the two-fermion bound states
behave effectively as $g=2$ exclusion particles, we end up with a mixture of $g=1$ and $g=2$ exclusion
statistics, where $\det (I - z H_{1,2})$ becomes a grand partition function  with $-z$
playing the role of the fugacity parameter. For example, for $q=5$
\bea
Z(4)
   =& & \tilde{s}_4 \tilde{s}_3 \tilde{s}_2 \tilde{s}_1 + \tilde{s}_5 \tilde{s}_3 \tilde{s}_2 \tilde{s}_1 + \tilde{s}_5 \tilde{s}_4 \tilde{s}_2 \tilde{s}_1 + \tilde{s}_5 \tilde{s}_4 \tilde{s}_3 \tilde{s}_1 + \tilde{s}_5 \tilde{s}_4 \tilde{s}_3 \tilde{s}_2 \nonumber\\
   &+& \tilde{s}_4 \tilde{s}_3 (-s_1) + \tilde{s}_5 \tilde{s}_3 (-s_1) + \tilde{s}_5 \tilde{s}_4 (-s_1)
+ \tilde{s}_4 \tilde{s}_1 (-s_2) + \tilde{s}_5 \tilde{s}_1 (-s_2) + \tilde{s}_5 \tilde{s}_4 (-s_2)  \nonumber\\
&+& \tilde{s}_2 \tilde{s}_1 (-s_3) + \tilde{s}_5 \tilde{s}_1 (-s_3) +\tilde{s}_5 \tilde{s}_2 (-s_3)
+ \tilde{s}_2 \tilde{s}_1 (-s_4) + \tilde{s}_3 \tilde{s}_1 (-s_4) + \tilde{s}_3 \tilde{s}_2 (-s_4) \nonumber\\
   &+& (-s_3) (-s_1) + (-s_4) (-s_1) + (-s_4) (-s_2) \nonumber
\eea
can be readily interpreted in figure \ref{figZ4} as the 4-body partition function for 4 particles, either
individual fermions or two-fermion bound states, occupying in all possible ways the five 1-body levels
$\epsilon_k,\; k=1, \dots, 5$.
Clearly when all $\tilde{s}_k$ are set to 0 in (\ref{1,2 det}), the $Z(2n\hp 1)$'s vanish and the $Z(2n)$'s reduce
to the $n$-body partition functions (\ref{g=2 Zn}) for $g=2$ exclusion particles, that is,
\vskip -0.6cm
\be \label{notbad}
\det(I-z H_{1,2})=\sum_{n=0,\, {\rm even}}^{q}(-z)^n Z(n)=\sum_{n=0}^{\lfloor q/2 \rfloor}z^{2n} Z(2n)
=\det(I-zH_2),
\ee
where in the last step we identified $(-1)^n Z(2n)$ to the $ Z(n)$ for $2$-exclusion appearing in $\det(I-zH_2)$  and given in (\ref{g=2 Zn}).
\begin{figure}[H]
\begin{center}
\includegraphics[scale=0.85]{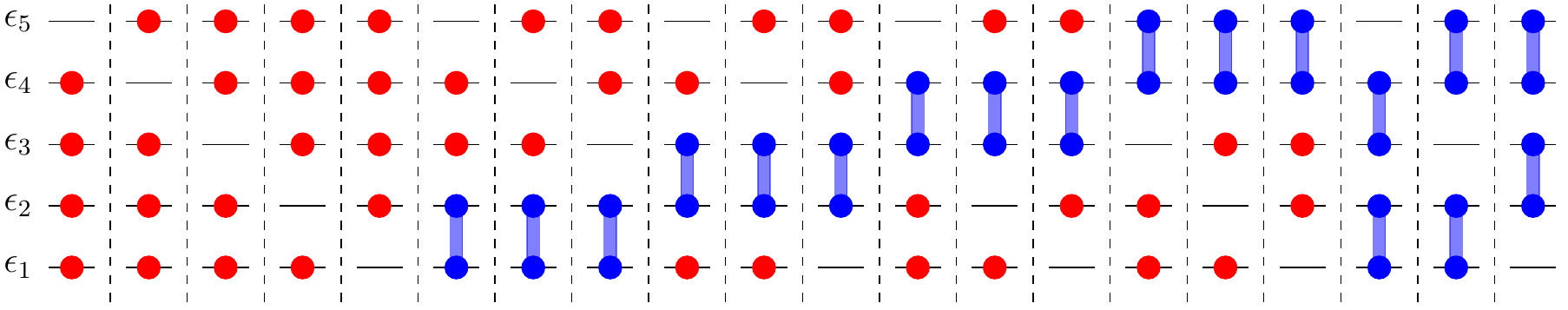}
\caption{\small{$Z(4)$ for $q=5$: all possible occupancies of the five $1$-body levels by 4 particles  with either fermions (red) or two-fermion bound states (blue).}}\label{figZ4}
\end{center}
\end{figure}
\noindent From 
\be \label{1,g Zn bn}\log\left( \sum_{n=0}^q Z(n)z^n\right)=\sum_{n=1}^{\infty}b(n)z^n
\ee implying
\be \nonumber
\tr \,H_{1,2}^{{\bf n}}={\bf n}(-1)^{{\bf n}+1}\hskip 0.05cm b({\bf n}),
\ee
one infers
\be \label{inferno}
b({\bf n})=(-1)^{{\bf n}+1}{\hskip -0.3cm}\sum_{\substack{\tilde{l}_1,\ldots,\tilde{l}_{j+1}; 
l_1 , \dots , l_j\\ (1,2)\text{-composition of }{\bf n}}}{\hskip -0.4cm} c_{1,2}(\tilde{l}_1,\sd,\tilde{l}_{j+1};
l_1,\sd ,l_j )\sum_{k=1}^{q-j} \tilde{s}_k^{\tilde{l}_1} s_k^{l_1} \tilde{s}_{k+1}^{\tilde{l}_2}{s}_{k+1}^{l_2}\cdots
\ee
with 
\bea \label{c 1,2}
c_{1,2}(\tilde{l}_1,\sd,\tilde{l}_{j+1}; l_1,\sd ,l_j)
&=&
\frac{{\tilde{l}_1+l_1 \choose l_1}}{\tilde{l}_1+l_1}~l_1\frac{{l_1+\tilde{l}_2+l_2\choose \tilde{l}_2\,,\,l_2}}{l_1+\tilde{l}_2+l_2}
\cdots l_{j}\frac{{l_{j}+\tilde{l}_{j+1}\choose l_{j}}}{l_{j}+\tilde{l}_{j+1}}
\\\nonumber
&=&\frac{(\tilde{l}_1+l_1\hm1)!}{\tilde{l}_1!~l_1!}
\prod_{k=2}^{j+1} { l_{k-1}+\tilde{l}_{k}+l_{k}\hm 1 \choose \tilde{l}_{k}~,~l_{k}}
\eea
with the usual convention $l_k = 0$ for $k>j$. 
We note that setting all $\tilde{l}_i$'s to zero reduces $c_{1,2}(\tilde{l}_1,\sd,\tilde{l}_{j+1}; l_1,\sd ,l_j)$ in 
(\ref{c 1,2}) to the standard 2-exclusion $c_{2}(l_1,\ldots,l_j)$ already discussed in (\ref{c2}).
Likewise, setting $\tilde{s}_k=0$ in (\ref{inferno}) eliminates all terms with nonzero $\tilde{l}_i$'s and
(\ref{c 1,2}) effectively reduces to (\ref{c2}).

We define the sequence of integers $\tilde{l}_1,\sd,\tilde{l}_{j+1};l_1,\sd,l_j$, $j\ge 0$, labeling 
$c_{1,2}$ in (\ref{c 1,2}) as a $(1,2)$-composition of the integer 
${\bf n}$ if they satisfy the defining conditions
\be
{\bf n}=(\tilde{l}_1+\tilde{l}_2+\cdots+ \tilde{l}_{j+1})+2(l_1+l_2+\cdots + l_j) 
~,~~ \tilde{l}_i \ge 0 ~,~~ l_i >0
\ee
That is, the $l_i$'s are the usual compositions of integers $1,2, \ldots, \lfloor {\bf n}/2 \rfloor$, while the ${\tilde l}_i$'s
are additional nonnegative integers (for $j\he 0$, we have the trivial composition
${\tilde l}_1 \hb = \hb {\bf n}$).
For example, there are six $(1,2)$-compositions of 4:
$(4)$, $(2,0;1)$, $(1,1;1)$, $(0,2;1)$, $(0,0;2)$, $(0,0,0;1,1)$, which contribute to $b(4)$ the terms
\be\nonumber
-b(4) = \frac{1}{4} \sum_{k=1}^q \tilde{s}_k^4 + \sum_{k=1}^{q-1} \tilde{s}_k^2 s_k + \sum_{k=1}^{q-1}  \tilde{s}_k s_k \tilde{s}_{k+1}+ \sum_{k=1}^{q-1} s_k \tilde{s}_{k+1}^2 + \frac{1}{2} \sum_{k=1}^{q-1} s_k^2 + \sum_{k=1}^{q-2} s_{k} s_{k+1}.
\ee
Note that the inverse of a composition, defined as 
$\tilde{l}_{j+1},\sd,\tilde{l}_1;l_j,\sd,l_1$ leaves $c_{1,2}$
invariant.

\subsection{ $(1,g)$-exclusion statistics}

For a mixture of $g=1$ and $g$ exclusion the associated algebraic area enumeration is encapsulated in the ($1,g$) exclusion matrix (again assuming zero ``umklapp" matrix elements at the off-diagonal corners)
\be\label{H 1,g}
H_{1,g}=\left(
\begin{array}{ccccccccc}
\tilde{s}_1 & f_{1} & 0 & \cdots & 0 & 0 & 0 & \cdots & 0 \\
0 & \tilde{s}_2 & f_{2} & \cdots & 0 & 0 & 0 & \cdots & 0 \\
\vdots & \vdots & \vdots & \ddots & \vdots & \vdots & \vdots & \ddots & \vdots \\
0 & 0 & 0 & \cdots & 0 & 0 & 0 & \cdots & 0 \\
g_{1} & 0 & 0 & \cdots & 0 & 0 & 0 & \cdots & 0 \\
0 & g_{2} & 0 & \cdots & 0 & 0 & 0 & \cdots & 0 \\
\vdots & \vdots & \vdots & \ddots & \vdots & \vdots & \vdots & \ddots & \vdots \\
0 & 0 & 0 & \cdots & 0 & 0 & 0 & \cdots & f_{q-1} \\
~0~ & ~0~ & ~0~ & \cdots & g_{q-g+1} & ~0~ & ~0~ & \cdots & \tilde{s}_q
\end{array}\right),
\ee
Following the same route as in the $g$-exclusion case, i.e., computing the secular determinant 
$\det(I-z H_{1,g})$, leads now to a mixture of fermions with Boltzmann factors 
${\rme}^{-\beta \epsilon_k} = {\tilde s}_k$ and
$g$-fermion bound states with $g$ particles occupying $g$ successive 1-body
levels $k,k\hp 1,\ldots,k\hp g\hm 1$ with Boltzmann factors
\be \label{s 1,g}
{\rme}^{-\beta \epsilon_{k,\dots,k+g-1}}=  (-1)^{g-1} s_k := (-1)^{g-1}\, g_{k}f_{k}f_{k+1}\cdots f_{k+g-2}
\ee
behaving effectively as $g$-exclusion particles. The associated cluster coefficients  are
\be \label{b 1,g}
b({\bf n})=(-1)^{{\bf n}+1} \hskip -0.3cm {\sum_{\substack{\tilde{l}_1,\ldots,\tilde{l}_{j+g-1};l_1,\dots,l_j\\ (1,g){\text{-composition of }}{\bf n}}}} \hskip -0.4cm 
c_{1,g}(\tilde{l}_1,\dots,{\tilde l}_{j+g-1}; l_1,\ldots,l_j)
\sum_{k=1}^{q-j-g+2} \tilde{s}_k^{\tilde{l}_1}s_k^{l_1}\tilde{s}^{\tilde{l}_2}_{k+1}s_{k+1}^{l_2}\cdots
\ee
We define the sequence of integers $\tilde{l}_1,\tilde{l}_2, \sd , \tilde{l}_{j+g-1}; l_1, l_2 , \sd , l_j$, $j\ge 1$,
as a $1,g$-composition of ${\bf n}$ if they satisfy the conditions
\bea
&{\bf n}=(\tilde{l}_1+\tilde{l}_2+\cdots + {\tilde l}_{j+g-1})+g(l_1+l_2+\cdots + l_j)  \nonumber\\
& {\tilde l}_i \ge 0~ ; ~ l_i \ge 0, ~  l_1,l_j >0 ,~ \text{at most}~ g\hm 2 ~\text{successive}~ \text{vanishing}~l_i 
\eea
That is, the $l_j$'s are the usual $g$-compositions of integers $1,2,\dots, \lfloor {\bf n}/g \rfloor$ and the ${\tilde l}_i$'s are additional nonnegative integers (we also include the trivial composition ${\tilde l}_1={\bf n}$.)
For example, there are seven $(1,3)$ compositions of $5$

\noindent
$\bullet$ $j=0$: $(5)$; $j=1$: $(2,0,0;1)$, $(1,1,0;1)$, $(1,0,1;1)$, $(0,2,0;1)$, $(0,1,1;1)$, $(0,0,2;1)$

\noindent and five $(1,4)$ compositions of $5$

\noindent
$\bullet$ $j=0$: $(5)$; $j=1$: $(1,0,0,0;1)$, $(0,1,0,0;1)$, $(0,0,1,0;1)$, $(0,0,0,1;1)$

\noindent The $c_{1,g}(\tilde{l}_1,\tilde{l}_2, \sd , \tilde{l}_{j+g-1}; l_1, l_2 , \sd , l_j)
$ in (\ref{b 1,g}) read
\bea \label{c g_1,g}
\hskip -0.5cm c_{1,g}(\tilde{l}_1,\tilde{l}_2, \sd , \tilde{l}_{j+g-1}; l_1, l_2 , \sd , l_j)&=&
\frac{(\tilde{l}_1\hp l_1\hm 1)!}{\tilde{l}_1!~l_1!}
\prod_{k=2}^{j+g-1}{{\tilde l}_k+\sum_{i=k-g+1}^k l_i \hm 1 \choose {\tilde l}_k ~ ,~ l_k} \nonumber
\eea
with $l_i =0$ for $i\le 0$ or $i>j$ as usual.
It is clear that when $\tilde{l}_i =0$ only the standard $g$-composition survives so that  the coefficients
$c_{1,g}$ in (\ref{c g_1,g}) go over to $c_g$ in (\ref{cg}). Equivalently, when ${\tilde s}_i=0$, terms
with non vanishing ${\tilde l}_i$ in (\ref{b 1,g}) drop and we recover the $g$-exclusion cluster coefficients.

Finally, we define the inverse of a composition by inverting the order of the ${\tilde l}_i$ and of
the $l_i$: ${\tilde l}_i \to {\tilde l}_{j+g-i} , l_i \to l_{j+1-i}$.
Inverse compositions produce the same coefficient $c_{1,g}$.

\subsection{Combinatorial interpretation}

$(1,g)$-compositions already have a combinatorial interpretation, deriving from their relation to cluster coefficients
of $(1,g)$-exclusion statistics. Specifically, $(1,g)$-compositions correspond to all {\it distinct connected}
arrangements of ${\bf n}$ particles on a 1-body spectrum, either alone or in $g$-bound states;
that is, to all the possible ways to place particles and bound states such that
they cannot be separated into two or more mutually non-overlapping groups (see figure \ref{figb5}).
If the arrangement covers $j\hp g\hm 1$ consecutive 1-body levels $k\hp i\hm 1$, 
$i = 1,\dots, j\hp g \hm 1$, then ${\tilde l}_i$ is the number of single particles on 1-body level $k\hp i \hm 1$
and $l_i$ is the number of $g$-bound states that 
extend over the $g$ levels $k\hp i \hm 1$ to $k\hp i \hp g\hm 2$ ($i = 1,\dots,j$).

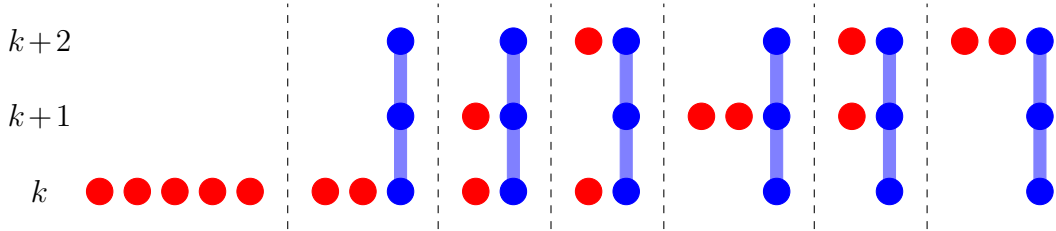
\begin{figure}[H]
\begin{center}
\begin{tikzpicture}[scale=1]
\draw[dashed] (2.5,-0.5) -- (2.5,2.5);
\draw[dashed] (4.5,-0.5) -- (4.5,2.5);
\draw[dashed] (6,-0.5) -- (6,2.5);
\draw[dashed] (7.5,-0.5) -- (7.5,2.5);
\draw[dashed] (9.5,-0.5) -- (9.5,2.5);
\draw[dashed] (11,-0.5) -- (11,2.5);
\node[align=center] at (-0.8,0) {$k$};
\node[align=center] at (-0.8,1) {$k\hp 1$};
\node[align=center] at (-0.8,2) {$k\hp 2$};
\filldraw[red] (0,0) circle (5pt);
\filldraw[red] (0.5,0) circle (5pt);
\filldraw[red] (1,0) circle (5pt);
\filldraw[red] (1.5,0) circle (5pt);
\filldraw[red] (2,0) circle (5pt);
\filldraw[blue,opacity=0.5] (4-0.08,0) rectangle (4+0.08,2);
\filldraw[blue] (4,0) circle (5pt);
\filldraw[blue] (4,1) circle (5pt);
\filldraw[blue] (4,2) circle (5pt);
\filldraw[red] (3,0) circle (5pt);
\filldraw[red] (3.5,0) circle (5pt);
\filldraw[blue,opacity=0.5] (5.5-0.08,0) rectangle (5.5+0.08,2);
\filldraw[blue] (5.5,0) circle (5pt);
\filldraw[blue] (5.5,1) circle (5pt);
\filldraw[blue] (5.5,2) circle (5pt);
\filldraw[red] (5,0) circle (5pt);
\filldraw[red] (5,1) circle (5pt);
\filldraw[blue,opacity=0.5] (7-0.08,0) rectangle (7+0.08,2);
\filldraw[blue] (7,0) circle (5pt);
\filldraw[blue] (7,1) circle (5pt);
\filldraw[blue] (7,2) circle (5pt);
\filldraw[red] (6.5,0) circle (5pt);
\filldraw[red] (6.5,2) circle (5pt);
\filldraw[blue,opacity=0.5] (9-0.08,0) rectangle (9+0.08,2);
\filldraw[blue] (9,0) circle (5pt);
\filldraw[blue] (9,1) circle (5pt);
\filldraw[blue] (9,2) circle (5pt);
\filldraw[red] (8,1) circle (5pt);
\filldraw[red] (8.5,1) circle (5pt);
\filldraw[blue,opacity=0.5] (10.5-0.08,0) rectangle (10.5+0.08,2);
\filldraw[blue] (10.5,0) circle (5pt);
\filldraw[blue] (10.5,1) circle (5pt);
\filldraw[blue] (10.5,2) circle (5pt);
\filldraw[red] (10,1) circle (5pt);
\filldraw[red] (10,2) circle (5pt);
\filldraw[blue,opacity=0.5] (12.5-0.08,0) rectangle (12.5+0.08,2);
\filldraw[blue] (12.5,0) circle (5pt);
\filldraw[blue] (12.5,1) circle (5pt);
\filldraw[blue] (12.5,2) circle (5pt);
\filldraw[red] (11.5,2) circle (5pt);
\filldraw[red] (12,2) circle (5pt);
\end{tikzpicture}
\end{center}
\caption{\small{Seven $(1,3)$ compositions of $5$: $(5)$, $(2,0,0;1)$, $(1,1,0;1)$, $(1,0,1;1)$, $(0,2,0;1)$, $(0,1,1;1)$, $(0,0,2;1)$, illustrated by fermions (red) and three-fermion bound states (blue).}}\label{figb5}
\end{figure}

Note that the above configurations are  forbidden by $(1,g)$-exclusion statistics. They constitute
the ``connected" components of the grand partition function, and their
exponentiation, with appropriate coefficients $(-1)^{{\bf n}+1} z^{\bf n} c_{1,g} ({\tilde l}_1,\sd;l_1,\sd )$,
produces the correct exclusion grand partition function, each term ``correcting" the overcounting arising
from the exponentiation of lower order terms. For pure fermion statistics $g=1$, all particles must occupy
the same level, leading to the trivial composition ${\tilde l}_1 = {\bf n}$ and the fermionic cluster
coefficients $(-1)^{{\bf n} +1} /{\bf n}$.

To give a combinatorial interpretation to the multiplicity coefficients
$c_{1,g} ({\tilde l}_1,\sd;l_1,\sd )$ we revert to the trace of
the ${\bf n}$-th power of $H_{1,g}$. In terms of the matrix elements $h_{ij}$ of $H_{1,g}^{\mathsf{T}}$
in (\ref{H 1,g}) this trace is
\bea \label{trace 1,g}
\tr\,H_{1,g}^{{\bf n}}
&=& \sum_{k_1=1}^q \sum_{k_2=1}^q \cdots \sum_{k_{\bf{n}}=1}^q h_{k_1 k_2}h_{k_2 k_3}\cdots h_{k_{\bf{n}} k_1}.
\eea
The structure of the $(1,g)$-exclusion matrix (\ref{Hg}) implies that (\ref{trace 1,g}) is a sum of products of 
${\bf n}$ factors $h_{k_i\hskip 0.015cm k_{i+1}}$ with indices such that $k_{i+1}\hm k_i$ take values $g\hm 1$, $0$, or $-1$.
We map the sequence of indices $k_1 , k_2 , \dots ,k_{{\bf n}-1}, k_{\bf n}, k_1$
to the heights of a periodic generalized $[g-1,0,-1]$ Motzkin path (``bridge") of length $\bf n$ starting and ending at height $k_1$,
with vertical steps up by $g-\hb 1$ units or down by $1$ unit as well as horizontal steps
(see figure \ref{Motzkin path example} for an example).
Evaluating the trace (\ref{direct}) amounts to
summing the corresponding products over all such periodic paths. We note that periodic paths must have
$g\hm 1$ down steps for each up step.
\begin{figure}[H]
\begin{center}
{\begin{tikzpicture}[scale=0.9]
\draw[thin, gray] (1,0) grid (13,3);
\draw[ultra thick, black](1,0)--(2,2)--(3,1)--(4,3)--(6,1)--(7,1)--(8,3)--(9,3)--(12,0)--(13,0);
\node[align=center] at (1,-0.5) {$1$};
\node[align=center] at (2,-0.5) {$2$};
\node[align=center] at (3,-0.5) {$3$};
\node[align=center] at (4,-0.5) {$4$};
\node[align=center] at (5,-0.5) {$5$};
\node[align=center] at (6,-0.5) {$6$};
\node[align=center] at (7,-0.5) {$7$};
\node[align=center] at (8,-0.5) {$8$};
\node[align=center] at (9,-0.5) {$9$};
\node[align=center] at (10,-0.5) {$10$};
\node[align=center] at (11,-0.5) {$11$};
\node[align=center] at (12,-0.5) {$12$};
\node[align=center] at (0.5,0) {$1$};
\node[align=center] at (0.5,1) {$2$};
\node[align=center] at (0.5,2) {$3$};
\node[align=center] at (0.5,3) {$4$};
\end{tikzpicture}}
\end{center}\vskip -0.5cm
\caption{\small {A periodic generalized Motzkin path of length ${\bf n}=12$ corresponding to the $(1,3)$-composition
$1,1,0,1;1,2$, starting with an up step from the first floor.}}\label{Motzkin path example}
\end{figure}
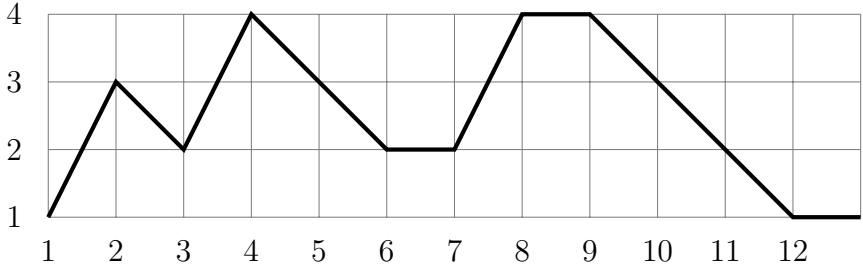

As in the  $g$-exclusion case, to group together terms with the same weight
$h_{k_1 k_2} \cdots h_{k_{\bf n} k_1}$ we need to consider paths with a fixed number of
transitions per level. For each path that reaches a lowest 1-body energy level $k$ and highest level from which
an up step starts
$k \hp j\hm 1$, and thus highest level reached $k\hp j \hp g\hm 2$, we denote by
$l_1,l_2,\ldots,l_j$ the number of up steps from levels
$k , k\hp 1, \dots, k\hp j\hm 1$ and by ${\tilde l}_1, {\tilde l}_2, \dots, {\tilde l}_{j+g-1}$
the number of horizontal steps at levels $k , k\hp 1, \dots,  k\hp j\hm g\hm 2$.
Clearly $\tilde{l}_1+\cdots + {\tilde l}_{j+g-1}+g(l_1+\cdots + l_j) \he {\bf n}$,
and at most $g\hm 2$ successive $l_i$ can vanish, since up steps can skip $g\hm 2$ floors.
Therefore, ${\tilde l}_1, \dots{\tilde l}_{j+g-1} ; l_1, \dots l_j$ is a $(1,g)$-composition of ${\bf n}$.
As before, each up step $k_i \to k_i\hp g\hm 1$ necessarily implies down steps 
$k_i \hp g\hm 1 \to k_i \hp g \hm2 ,\dots, k_i\hp 1 \to k_i$,
so factors in each term in (\ref{trace 1,g}) corresponding to each up step
$k_i \to k_i\hp g\hm 1$ contribute the full combination 
$h_{k_i ,k_i +g-1} \, h_{k_i +g-1, k_i +g-2}  \cdots h_{k_i +1,k_i }= g_{k_i} \, f_{k_i +g-2} \cdots f_{k_i} 
= s_{k_i}$.

Altogether, the sum in (\ref{trace 1,g}) rewrites  as
\be
\tr\,H_{1,g}^{\bf n}=\sum_{k=1}^{q-j-g+2} {\sum_{\substack{\tilde{l}_1,\ldots,\tilde{l}_{j+g-1};l_1,\dots,l_j\\ (1,g){\text{-composition of }}{\bf n}}}} 
C_{1,g}(\tilde{l}_1 ,\sd, \tilde{l}_{j+g-1}; l_1, \sd , l_j) \,
\tilde{s}_k^{\tilde{l}_1}s_k^{l_1}\tilde{s}^{\tilde{l}_2}_{k+1}s_{k+1}^{l_2}\cdots
\nonumber\ee
where $C_{1,g} ({\tilde l}_1, \sd ; l_1,\sd )$ is the number of periodic generalized Motzkin paths of length
${\bf n}$ with $\tilde{l}_1$ horizontal steps and $l_1$ up steps originating from the first floor, $\tilde{l}_2$
and $l_2$ from the second floor, etc.~(the sum
over $k$ ensures that paths of all possible starting level $k_1$ in (\ref{trace 1,g}) are included). Comparing with
\be \label{tr 1,g}
\tr\,H^{{\bf n}}_{1,g}= {\sum_{\substack{\tilde{l}_1,\ldots,\tilde{l}_{j+g-1};l_1,\dots,l_j\\ (1,g){\text{-composition of }}{\bf n}}}} \hskip -0.4cm 
c_{1,g}(\tilde{l}_1,\dots,{\tilde l}_{j+g-1}; l_1,\ldots,l_j)
\sum_{k=1}^{q-j-g+2} \tilde{s}_k^{\tilde{l}_1}s_k^{l_1}\tilde{s}^{\tilde{l}_2}_{k+1}s_{k+1}^{l_2}\cdots
\nonumber\ee
we see that
\be
C_{1,g}(\tilde{l}_1 ,\sd, \tilde{l}_{j+g-1}; l_1, \sd , l_j) = {\bf n}\, c_{1,g}
(\tilde{l}_1 ,\sd, \tilde{l}_{j+g-1}; l_1, \sd , l_j).
\nonumber\ee
Therefore, ${\bf n}\, c_{1,g}
(\tilde{l}_1 ,\sd, \tilde{l}_{j+g-1}; l_1, \sd , l_j)$ admits the interpretation of the
number of periodic generalized [$g\hm 1,0,-1$]  Motzkin paths with horizontal and up steps as defined above.

The number of such paths starting with an up step, resp. a horizontal step, from
floor $i$ can also be deduced as $l_i\, c_{1,g} (\tilde{l}_1 ,\sd, \tilde{l}_{j+g-1}; l_1, \sd , l_j)$,
resp. ${\tilde l}_i\, c_{1,g} (\tilde{l}_1 ,\sd, \tilde{l}_{j+g-1}; l_1, \sd , l_j)$, while the total number
of paths starting from floor $i$ is 
\be\nonumber
\left({\tilde l}_i +\sum_{k=i-g+1}^i l_k\right) c_{1,g} (\tilde{l}_1 ,\sd, \tilde{l}_{j+g-1}; l_1, \sd , l_j).
\ee
(the result $({\tilde l}_1 + l_1)\, c_{1,2}$ for $i\he1$ Motzkin excursions was also derived in \cite{Motzkin}).
Finally, the number of paths starting at floor $i$ with a down step can be deduced as
\be\nonumber
\left(\sum_{k=i-g+1}^{i-1} l_k\right) c_{1,g} (\tilde{l}_1 ,\sd, \tilde{l}_{j+g-1}; l_1, \sd , l_j).
\ee
%

The proof of the above counting formulae can be obtained similarly to the case of $g$-exclusion and
generalized Dyck paths. The simplest method is the one outlined at the end of section \ref{Geng} based on
cyclic permutations. Consider first
paths that start with an up step from the first floor. A combinatorial argument entirely analogous to the one
in section \ref{Geng} yields the result $l_1\, c_{1,g}$ for the number of such paths, and by periodic permutation and
reduction by an overcounting factor of $l_1$
the total number of paths obtains as ${\bf n} \, c_{1,g}$. A repetition of the periodic argument from
floor $i$, then, produces the results $l_i\, c_{1,g}$ and ${\tilde l}_i \, c_{1,g}$ for the number of
paths starting up or horizontally from floor $i$, and a `cutting and exchanging' argument gives the
number of paths starting down from floor $i$. The details are similar to the ones in section
\ref{Geng} and are left as an exercise.

We conclude by giving the number of $(1,g)$-compositions of  a given integer $\bf n$
\be \label{Nn}
{N}_{1,g} ({\bf n}) = 1+\sum_{k=0}^{\lfloor {\bf n}/g \rfloor -1} \sum_{m=0}^{(g-1)k} 
\binom{k}{m}_{\bb\bb g} \binom{{\bf n}+m-gk-1}{m+g-1}\;,
\ee
where the $g$-nomial coefficient is defined as
\bea \nonumber
\binom{k}{m}_{\bb\bb g}&=&[x^m](1+x+x^2+\cdots+x^{g-1})^k = [x^m] \left({1-x^g \over 1-x}\right)^k\\\nonumber
&=& \sum_{j=0}^{\lfloor m/g \rfloor}(-1)^j \binom{k}{j} \binom{k+m-gj-1}{k-1}.
\eea
For $g\he2$ it reduces to the standard binomial coefficient $\binom{k}{m}_2=\binom{k}{m}$.
So (\ref{Nn}) becomes the triple sum 
\be \nonumber
{N}_{1,g} ({\bf n}) = 1+\sum_{k=0}^{\lfloor {\bf n}/g \rfloor -1} \sum_{m=0}^{(g-1)k} \sum_{j=0}^{\lfloor m/g \rfloor}(-1)^j \binom{k}{j} \binom{k+m-gj-1}{k-1}
\binom{{\bf n}+m-gk-1}{m+g-1}.
\ee
Equivalently, the generating function of the $N_{1,g} ({\bf n})$'s is
\bea
 \sum_{{\bf n}=0}^\infty x^n \, {N}_{1,g} ({\bf n}) &=& 
{(1\hm x)^{g-2} (1\hp x^{g-1}\hm x^g)-x^{g-1} \over (1\hm x)^{g-1} (1\hp x^{g-1}-x^g)\hm x^{g-1}} \\
&=& {1\over 1\hm x} \left[1+ {x^g \over (1\hm x)^{g-1} (1\hp x^{g-1}\hm x^g)-x^{g-1} }\right].\nonumber
\eea
In the second line, the term $1/(1-x)$ reproduces the trivial compositions (${\tilde l}_1 = {\bf n}$)
while the other term reproduces all the non trivial ones.
Finally, the number of all unrestricted periodic $[g\hm 1, 0, -1]$ generalized Motzkin paths 
is obtained by summing $c_{1,g}$ over all $1,g$-compositions and yields the relation
\be \nonumber
{\bf n}\hskip -0.3cm{\sum_{\substack{\tilde{l}_1,\ldots,\tilde{l}_{j+g-1};l_1,\dots,l_j\\ (1,g){\text{-composition of }}{\bf n}}}} \hskip -0.5cm c_{1,g}(\tilde{l}_1 ,\sd, \tilde{l}_{j+g-1}; l_1, \sd , l_j)=[x^0](x^{g-1}+1+x^{-1})^{{\bf n}} =
\sum_{k=0}^{\lfloor {\bf n}/g \rfloor}{{\bf n}\choose gk}{gk\choose k}.
\ee

\section{Conclusions}
We have established a connection between the enumeration of lattice walks according to their algebraic area,
quantum exclusion statistics, and the combinatorics of generalized Dyck and Motzkin paths (also known as
{\L}ukasiewicz paths). The key common quantities are the coefficients $c_g(l_1,\ldots,l_j)$ and
$c_{1,g}(\tilde{l}_1,\ldots,\tilde{l}_{j+g-1};l_1,\ldots,l_j)$
labeled by the $g$-compositions and the $(1,g)$-compositions of the length of the walks. These coefficients
appear as essential building blocks of the algebraic area partition function of walks on square or honeycomb lattices, the cluster coefficients $g$-exclusion and $(1,g)$-exclusion statistical systems, and the counting
of generalized paths with specific number of steps from each visited floor. The connection of Dyck paths
and $g=2$ exclusion statistics was established in \cite{dyke} and used to calculate the length and area
generating function for such paths, and the method was extended to Motzkin paths in \cite{motzi}.
To the best of our knowledge, the full threefold connection between walks, statistics, and paths, as well as
the explicit expressions of $c_g$ and $c_{1,g}$ for $g>2$ and their relevance to {\L}ukasiewicz
path counting, were put forward for the first time in the present work.

There are various directions for possible future investigation. The most immediate one is along the lines
already laid out in this work, that is, in the connection of walks of various properties and on various lattices
and corresponding paths. For instance, the enumeration of open walks on the square lattice according to their
algebraic area was recently achieved \cite{open}, with Dyck path combinatorics again playing a key role.
Walks on other lattices, such as the kagom\'e lattice, and of different properties can be investigated
with similar methods.

The concept of exclusion statistics and related
compositions naturally generalizes to $(g', g)$ and more general $(g_1, g_2, \dots, g_n)$ statistics and
compositions, and the statistical mechanical properties of these systems and mathematical properties of their
compositions are of interest. It would also be worthwhile to derive the corresponding combinatorial quantities
$c_{g_1 , \dots , g_n}$ and study their relevance for generalized {\L}ukasiewicz paths.

In a different direction, it is known that Dyck and Motzkin paths appear in various contexts in
physics and mathematics. In physics, they appear in percolation processes, interfaces
between fluids of different surface tension, and other statistical systems (see, e.g., \cite{Prell}).
Further, they can be mapped to
spin-1/2 and spin-1 chains. E.g., in \cite{Bravyi} a spin-1 frustration-free Hamiltonian was constructed using
Motzkin paths. In a related direction, a family of  multispin quantum chains with a free-(para)fermionic
eigenspectrum \cite{baxter} was recently reanalyzed  in \cite{Alcaraz} and the eigenenergies were obtained
via the roots of a polynomial with coefficients similar to the $Z(n)$ in the present paper, indicating a connection
with exclusion statistics that warrants further investigation. Finally, in knot theory, the Temperley–Lieb algebra
can have a representation based on Dyck paths, while, if empty vertices (vertices not incident to an edge) are
allowed, Motzkin paths become relevant \cite{Motzkin algebras}. The extension of this connection to more
general paths, and the meaning of the $c_g$ and $c_{1,g}$ coefficients in this context, are nontrivial issues
that deserve further study.

\vskip 0.5cm

\noindent
{\bf Acknowledgments}

\noindent
L.G. acknowledges the financial support of China Scholarship Council (No. 202009110129). The work of A.P. was supported in part by NSF under grant NSF-PHY-2112729 and by a PSC-CUNY grant.

\end{document}